\documentclass[showpacs,preprintnumbers,amsmath,amssymb,nofootinbib]{revtex4}

 \usepackage{graphics}
 \usepackage{amsmath}
 \usepackage{amsfonts}
 \usepackage{amssymb}
 \usepackage{epsfig}
 \usepackage{graphicx}
 \usepackage{color}

 \allowdisplaybreaks[4]

 \newcommand{\eqlabel}[1]{
   \label{#1}
 }
 \newcommand{\seclabel}[1]{
   \label{#1}
 }
 \newcommand{\figlabel}[1]{
   \label{#1}
 }

 \newcommand{\nn}{\nonumber}
 \newcommand{\er}[1]{(\ref{#1})}          % Equation reference with brackets

 \renewcommand{\d}{\mathrm d}  % total derivative d
 \newcommand{\td}[2]{\frac{{\rm d} #1}{{\rm d} #2}}     % total derivative fraction
 \newcommand{\tdil}[2]{{\rm d} #1/{\rm d} #2}     % total derivative in line
 \newcommand{\pdil}[2]{\partial #1/\partial #2 }     % partial derivative in line
 \newcommand{\tdt}[2]{ \frac{{\rm d}^2 #1}{{\rm d} {#2}^2} }     % second total derivative fraction
 \newcommand{\tdtil}[2]{ {\rm d}^2 #1/{\rm d} {#2}^2 }     % second total derivative in line
 \newcommand{\Rt}{\dot{R}}
 \newcommand{\Rh}{\hat{R}}
 
 \newcommand{\Rrh}{\widehat{R'}}
 
 \renewcommand{\th}{\hat{t}}
 \newcommand{\rh}{\hat{r}}
 \newcommand{\rhoh}{\hat{\rho}}

\begin{document}

 \title{Determining the metric of the Cosmos: stability, accuracy, and 
consistency}

 \author{M. L. McClure}
 \email{mcclure@astro.utoronto.ca}
 \author{Charles Hellaby}
 \email{Charles.Hellaby@uct.ac.za}
 \affiliation{Department of Mathematics \& Applied Mathematics, 
University of Cape Town, Rondebosch 7701, South Africa}
 
 \date{March 28, 2008}

 \begin{abstract}

The ultimate application of Einstein's field equations is to empirically
determine the geometry of the Universe from its matter content, rather
than simply assuming the Universe can be represented by a homogeneous
model on all scales.  Choosing an LTB model as the most convenient
inhomogeneous model for the early stages of development, a data reduction
procedure was recently validated using perfect test data.  Here we
simulate observational uncertainties and improve the previous numerical
scheme to ensure that it will be usable with real data as soon as
observational surveys are sufficiently deep and complete.  Two regions
require special treatment---the origin and the maximum in the areal
radius.  To minimize numerical errors near the origin, we use an LTB
series expansion to provide the initial values for integrating the
differential equations.  We also use an improved method to match the
numerical integration to the series expansion that bridges the region near
the maximum in the areal radius. Because the mass enclosed within the
maximum obeys a specific relationship, we show that it is possible to
correct for a fixed systematic error in either the distance scale or the
redshift-space mass density, such that the integrated values are
consistent with the data at the maximum. 

 \end{abstract}


 \maketitle

 \section{Introduction}

   Einstein's equations relate the geometry of spacetime to the 
distribution of matter, and the most important application of this key 
physical concept is the study of the Universe.  
As the amount and precision of cosmological data improve, we will be able 
to determine not only the overall curvature of the Universe, but its 
detailed geometric structure too.

It is generally assumed that the Universe is homogeneous above 
the scale of superclusters and that it can be represented by a 
Friedmann-Lema\^{i}tre-Robertson-Walker (FLRW) model. 
This assumption has served cosmology well and led to a very good 
understanding of many features of the observed Universe, which 
implies it cannot be seriously wrong.
Yet it must be acknowledged that homogeneity is indeed an assumption 
that should eventually be checked against observations, once the data 
are sufficiently accurate and complete over a large enough range of 
redshifts.  
Because of our fixed position in the Universe, there are two distinct 
aspects to homogeneity---isotropy about our position, and 
uniformity with distance from us.  
Isotropy is relatively easy to check, but radial homogeneity has 
many intrinsic complications.  
Therefore a thorough verification will require considerable effort.
The assumption of homogeneity is so pervasive, and underlies so much 
theoretical and observational work, that there is a real danger of a 
circular argument.  
Consequently, any proof of homogeneity must ensure it does not rely on 
results obtained using an assumption of homogeneity.
Checking that the distribution of matter is homogeneous, while still 
using an FLRW model, is at best a consistency check and not a proof.  
A genuine proof must allow both the geometry and the matter to be 
inhomogeneous.  

Historically, it has been theorized that it should be possible to use 
astronomical measurements to directly determine the metric of the Cosmos.
The idea of co-ordinates based on the observer's past null cone was 
first suggested by Temple \cite{Tem38}, who called them ``optical 
co-ordinates."  
McCrea surveyed observational relations in homogeneous models \cite{McC34} 
and was the first to determine them, as a series approximation, for 
inhomogeneous models \cite{McC38}.
Kristian and Sachs \cite{Kri66} were the first to consider the 
determination of a general metric for the Universe from observational 
data.  
They used series expansions near the observer in powers of diameter 
distance (and density field) for redshift, image distortion, number 
density, and proper motion.  
They considered the case of dust and estimated the parameters in their 
series, concluding that homogeneity was not proven.
Ellis et al.\ \cite{Ell85} provided a major review of this problem.  
They showed it is not possible to obtain the spacetime metric without 
assuming Einstein's field equations, and they also raised the
problem of source evolution.  
Further papers by Stoeger, Nel, Maartens, Araujo, and others
\cite{SNME92,SEN92,SNE92b,SNE92c,MaartensMatravers,MHMS96,AS99,AABFS01,ARS01}
considered the fluid-ray tetrad, the spherically symmetric case and its
perturbations, various solutions, and the origin conditions in these 
co-ordinates.
Mustapha et al.\ \cite{MBHE} showed the diameter distance and the density 
observed on the past null cone can be significantly distorted by inhomogeneity.
Ribeiro and Stoeger \cite{RS03} considered the inclusion of a galaxy 
luminosity function, and checked catalogue data for consistency with the 
theory.  A follow-up paper of Albani et al.\ \cite{AlIrRiSt07} 
showed that analyses of galaxy 
statistics are strongly affected by the distance definition used.

Mustapha, Hellaby, and Ellis \cite{Mus97} considered data that are 
isotropic about us, and described an algorithm that could determine the 
specific Lema\^{i}tre-Tolman-Bondi (LTB) model \cite{Lem33,Tol34,Bon47}
that reproduces the observations.  
Thus, they claimed that any reasonable sets of observations of diameter 
or luminosity distance against redshift, and number counts against 
redshift, can be fit by an LTB model.  
They also showed that the effects of source evolution, cosmic evolution, 
and radial inhomogeneity are mixed together on the past null cone, and 
it is difficult to distinguish them.  However, Hellaby \cite{Hel01} 
suggested a method of checking theories of source evolution in 
inhomogeneous geometries using multi-color observations.  
C\'el\'erier \cite{Cel00} considered a series expansion on the past null 
cone of an LTB model, and showed that the supernova data can be reproduced.
Recently C\'el\'erier reviewed \cite{Cel07a,Cel07b} the need for 
inhomogeneous models in relation to observations, concluding that the 
standard methods of fitting a homogeneous universe to the data are open 
to significant systematic errors, and noting that fitting observations 
with inhomogeneous models is a significant challenge.  
Bishop and Haines \cite{Bish96} approached the Kristian and Sachs program 
as a time-reversed characteristic initial value problem, but they didn't 
solve the problem of the maximum in the diameter distance, and they only 
tested their numerical code with Einstein-de Sitter data.

Nevertheless, other than very limited series expansions, the 
various proposed methods have never been implemented with real data.
Therefore we have begun a project to develop a numerical procedure that 
can do the necessary calculations and handle observational data.  
Thus, when sufficient data become available in the 
not-too-distant future, we will 
have a numerical routine capable of determining the Universe's spacetime 
structure.

The algorithm suggested by Mustapha, Hellaby, and Ellis \cite{Mus97} was 
implemented as a numerical procedure by Lu and Hellaby \cite{Lu06,Lu07}.  
The process of turning the algorithm into code raised some important 
issues that had not been obvious when discussing the theory.  
These include the fact that the real data are discrete, there are no data 
at the origin itself, and the differential equations to be 
solved are singular at the maximum in the areal radius, as well as the
problem of the choice of appropriate numerical methods, given that the 
differential equations contain both observational data and the functions 
being solved for.
The program was tested using ideal data, for a variety of homogeneous 
and inhomogeneous models, and successfully recovered the original 
models, thus demonstrating the viability of the method.

The main reason for working with spherical symmetry in these initial 
stages is merely to keep the theory and the numerics relatively simple.  
In the long run, this assumption will be dropped.  
However, there are three observational reasons why spherical symmetry is 
a good place to start.  
First, we note that we are at the center of our past null cone, so it makes 
sense to consider spacetime in terms of spherical co-ordinates.
Second, the Universe does appear nearly isotropic on large scales, but 
radial homogeneity is not easy to verify because of the finite travel 
time of light and the miniscule duration over which cosmological 
observations have been made, so it is more urgent to determine the 
radial variation of the metric.
Finally, there is no deep all-sky redshift survey at present, and the 
zone of avoidance is likely to be a gap in any survey for the
foreseeable future.

If observations are made over the whole sky and the Universe is truly 
homogeneous on large scales, then using a spherically-symmetric model and 
binning the data over all angles should smear out any structures and 
lead to a radially-homogeneous metric at radii larger than the largest 
structures.  
At smaller radii the spherical shells may be enclosed entirely within a 
void or dominated by a supercluster and would appear inhomogeneous.  
Also, as a first check on isotropy, measurements made in different 
directions on the sky could be reduced separately, using our 
spherically-symmetric procedure, to look for angular variation.

The aim of the present paper is to extend the previous numerical routine 
of Lu and Hellaby \cite{Lu07} to be able to handle observational 
uncertainties.  
We study the stability of the differential equations used in the 
numerical routine, and how the uncertainties in the data should be 
propagated through the integration of the differential equations.  
We 
approximate the effect of random errors in observational data by adding 
Gaussian deviates to test data generated from model universes.  
Using this simulated data, we develop various impovements to the original 
numerical procedure, especially for calculating the derivatives of the 
areal radius, for dealing with the regions near the origin, and for 
matching the series expansion used about the maximum in the areal radius 
with the numerical integration regions on either side.
Finally, we study the effect of systematic errors on the numerical 
algorithm and show that it is possible to correct for an overall 
systematic error in the distance scale or in the density (or if both are 
present, one may be corrected relative to the other to obtain a 
self-consistent result).  
Hellaby \cite{Hel06} showed that there is a relationship between the 
enclosed gravitational mass $M$, the areal radius $R$, and the 
cosmological constant $\Lambda$ at the maximum in $R$ that is independent 
of any inhomogeneity:
this relationship is the key to detecting and correcting for systematic 
errors.

The main concern in this paper is the development of a basic numerical 
procedure that can handle input data containing statistical and 
systematic errors.  
Of course, considerable processing is needed to extract the data 
functions needed by our numerical algorithm from the raw astronomical 
observations---luminosity functions, K-corrections, evolution of galaxies 
in luminosity and mass, galaxy mergers, etc.---and many sources of error 
must be considered in assessing the uncertainties.  
Though these must be carefully addressed in the future, for the present 
we assume that we are given data in the appropriate form with known 
uncertainties.
It seems that type Ia supernovae are establishing themselves as very 
reliable standard candles out to redshifts well above 1, and when a 
sufficient number are measured, they may well provide the luminosity 
distance data for this project, or the basic calibration for a range of 
sources.
Determination of the density of matter in redshift space, which is 
composed of the number density of sources and the mass per source, is 
much more problematic.  
The determination of masses from dynamics does extract the desired 
gravitational masses, but requires detailed observations, is not easy to 
do in large numbers, and can only be performed where there is luminous 
matter.  It would be very convenient if there were a 
well-established tracer of mass that could be used instead of trying to 
count everything.  
There are some suggestions, such as luminous red galaxies, but it is very 
hard to demonstrate a certain type of source is a reliable tracer until a 
full mass census is taken over a region including several structures of 
all sizes.  
If the tracer sources are too widely dispersed, this would limit the 
scale on which inhomogeneity can be detected.
Although current cosmological data are well short of the required quality 
and quantity, we expect dramatic improvements in accuracy and 
completeness in the not-too-distant future.

\section{The Model and the Null Cone}

We use the spherically-symmetric inhomogeneous-dust model (the ``LTB" model) 
discovered by Lema\^{\i}tre, rediscovered by Tolman, and studied by Bondi 
\cite{Lem33,Tol34,Bon47}.
We adhere to the notation used by Lu and Hellaby \cite{Lu07}.
The metric is
 \begin{align}
   \d s^2 = - \d t^2 + \frac{(R')^2}{1 + 2 E} \, \d r^2 + R^2 \d \Omega^2 ~,
   \eqlabel{ds2LT}
 \end{align}
 where $\mathrm{d}\Omega^2 = \mathrm{d}\theta^2 + \sin^2\theta \, \mathrm{d} \phi^2$, $R(t,r)$ is the areal radius, 
a prime denotes partial differentiation with respect to $r$, 
and the free function $E(r) \geq -1/2$ is a localized geometry 
term.  From the Einstein field equations we get
 \begin{align}
   \Rt^{2} = \frac{2 M(r)}{R} + 2 E(r) + \frac{\Lambda R^2}{3} ~, \eqlabel{RtSq}
 \end{align}
 where $\Rt = \pdil{R}{t}$, and
 \begin{align}
   8 \pi \rho = \frac{2 M'}{R^2 R'} ~, \eqlabel{rho}
 \end{align}
 where $M(r)$ is a second free function that gives the gravitational mass within a 
comoving shell of radius $r$.  
Here $E(r)$ also plays the role of the local
energy per unit mass of the dust particles.  For the present we take $\Lambda = 0$, 
postponing more general considerations for later work.  The solutions of \er{RtSq},
in terms of parameter $\eta$, are
 \begin{align}
   E > 0:~~~~~~   R & = \frac{M}{2 E} \, (\cosh \eta - 1) ~,~~~~~~ &
      (\sinh \eta - \eta) & = \frac{(2 E)^{3/2} (t - t_B)}{M} ~;
      \eqlabel{HypEv} \\
   E = 0:~~~~~~   R & = M \left( \frac{\eta^2}{2} \right) ~,~~~~~~ &
      \left( \frac{\eta^3}{6} \right) & = \frac{(t - t_B)}{M} ~;
      \eqlabel{ParEv} \\
   E < 0:~~~~~~   R & = \frac{M}{(-2 E)} \, (1 - \cos \eta) ~,~~~~~~ &
      (\eta - \sin \eta) &= \frac{(-2E)^{3/2} (t - t_B)}{M} ~;
      \eqlabel{EllEv}
 \end{align}
 for hyperbolic, parabolic, and elliptic evolution respectively.  (Near the origin, 
where $E \to 0$, the type of evolution is determined by the sign of $RE/M$ or 
$E/M^{2/3}$.)  These solutions contain a third free function $t_B(r)$, which is the 
time of the big bang locally.  By specifying the three free functions---$M(r)$, $E(r)$, 
and $t_B(r)$---an LTB model is fully determined.  Between them they provide a radial 
co-ordinate freedom and two physical relationships.

 \subsection{The observables and the differential equations}

The background theory is presented in Section 2 of Lu and Hellaby 
\cite{Lu07}, and here we merely summarize the essentials.

 The diameter and luminosity
 \footnote{In Kristian and Sachs \cite{Kri66} a ``corrected luminosity 
distance" was defined to be the same 
as the diameter distance.  Some authors have called this the ``luminosity 
distance," which has led to a confusion of terminology and sometimes to 
incorrect definitions.}
 distances are
 \begin{align}
   \Rh = d_D = \frac{D}{\delta} ~,~~~~~~~~
   d_L = \sqrt{\frac{L}{\ell}}\; \, d_{10} ~,
   \eqlabel{dA-dL}
 \end{align}
 where $D(z)$ and $L(z)$ are the true diameter and absolute luminosity of 
a source, $\delta$ and $\ell$ are the corresponding angular diameter and 
apparent luminosity, $d_{10}$ is $10$~parsecs, and they are related by 
the reciprocity theorem \cite{Ellis71}
 \footnote{
 This was first shown by Etherington \cite{Eth33}.  
 Kristian and Sachs \cite{Kri66} cite a private communication from R. 
Penrose for a general proof of the reciprocity theorem, but we are not 
aware of any publication of that work.} 
 \begin{align}
   (1 + z)^2 d_D = d_L ~.   \eqlabel{dDdL}
 \end{align}
 The redshift-space number density of sources $n$, measured in number per 
steradian per unit redshift interval, is related to the density $\rho$ by 
(\cite{Mus97,Lu07})
 \begin{align}
   {\Rh}^2 \rhoh = \mu n \td{z}{r} ~,
   \eqlabel{rho-n}
 \end{align}
 where $\mu$ is the mass per source.

   Light rays arriving at the central observer follow $\mathrm{d}s^2 = 0 
= \mathrm{d} \theta^2 = \mathrm{d} \phi^2$, 
so the past null cone of the observation event ($t = t_0, r = 0$) satisfies
 \begin{align}
   \td{t}{r} = - \frac{R'}{W} ~,~~~~~~~~ W = \sqrt{1 + 2 E} ~,
   \eqlabel{dtdrNC}
 \end{align}
 and we write the solution $t = \th(r)$ or $r = \rh(t)$.  
We denote a quantity evaluated on the observer's past null cone with a hat on top or as 
a subscript, for example $R(\th(r),r) \equiv \Rh$ or $[R]_\wedge$, though this will often 
be omitted where it is obvious from the context.  

   We can use the radial co-ordinate freedom to choose
 \begin{align}
   \td{\th}{r} = -1 ~,~~~~\mbox{i.e.}~~~~ \Rrh = W ~,
   \eqlabel{RrPNC}
 \end{align}
 on the observer's past null cone, so that the solution to \er{dtdrNC} is
 \begin{align}
   \th = t_0 - r ~.
   \eqlabel{rPNC}
 \end{align}
 Note that \er{RrPNC} and \er{rPNC} and the following differential equations 
only hold for the single null cone with apex $(t_0, 0)$.  

 Since the co-ordinate $r$ is not an observable, all $r$ derivatives are 
converted to $z$ derivatives, defining
 \begin{align}
   \phi = \td{r}{z} = - \frac{\td{\Rh}{r}}
{(1 + z) \left( 4 \pi \mu n \td{z}{r} + \tdt{\Rh}{r} \right)}
   \eqlabel{phiDef}
 \end{align}
 and using
 \begin{align}
   \td{\Rh}{z} = \td{\Rh}{r} \, \phi ~,~~~~~~~~
   \tdt{\Rh}{z} = \tdt{\Rh}{r} \, \phi^2 + \td{\Rh}{r} \, \td{\phi}{z} ~.
   \eqlabel{dRhdzChain}
 \end{align}
 Then from Lu and Hellaby \cite{Lu07}, the differential equations are
 \begin{align}
   \td{M}{z} & = 4 \pi \mu n W ~,
   \eqlabel{dMdz} \\
   W & = \frac{1}{2 \phi} \left( \td{\Rh}{z} \right) + 
   \frac{\left( 1 - \frac{2 M}{\Rh} \right) \phi}{2 \left( \td{\Rh}{z} \right)}
   \eqlabel{WPNCz}
 \end{align}
and
 \begin{align}
   \td{\phi}{z} = \phi \left( \frac{1}{(1 + z)}
   + \frac{\frac{4 \pi \mu n \phi}{\Rh} + \tdt{\Rh}{z}}{\td{\Rh}{z}} \right) ~.
   \eqlabel{dphidz}
 \end{align}

 Equations \er{phiDef}, \er{dphidz}, \er{dMdz}, and \er{WPNCz} constitute the 
differential equations (DEs)
to be solved for $\phi(z)$, $r(z)$, $M(z)$, and $E(z)$.  Then $t_B(z)$ 
follows from \er{HypEv}-\er{EllEv} 
and \er{rPNC}.  
Knowing $r(z)$, $M(z)$, $E(z)$, and $t_B(z)$ fully determines the LTB metric that 
reproduces the given $\Rh(z)$ and $\mu(z) n(z)$ data.  

   The LTB origin conditions were thoroughly examined in Mustapha and 
Hellaby \cite{MH01}, and Lu and Hellaby \cite{Lu07} give their 
application in this case.  

 \subsection{Apparent horizon}
 \seclabel{ApHor}

   The locus where the past null cone crosses the apparent horizon is also where
the areal radius is maximum, i.e.\ the maximum $\Rh$ is $\Rh = R_m$ at $z = z_m$.
Since $\tdil{\Rh}{z} = 0$ here, the DEs \er{dphidz} and \er{dMdz} with \er{WPNCz}
become singular.  
However, it is obvious from \er{WPNCz} that where $\tdil{\Rh}{z} = 0$ we must 
also have
 \begin{align}
   \Rh_m = 2 M_m ~,   \eqlabel{RhmEq}
 \end{align}
 since $W$ is arbitrary (see Krasi\'{n}ski and Hellaby \cite{KraHe04}, 
Hellaby \cite{H87}).  Similarly \er{phiDef} and \er{dphidz} show that 
 \begin{align}
   \tdt{\Rh}{r} \Bigg|_m \phi_m = - 4 \pi \mu_m n_m ~,~~~~~~~~
   \tdt{\Rh}{z} \Bigg|_m = - 4 \pi \mu_m n_m \phi_m ~,   \eqlabel{Rhzzm}
 \end{align}
 since we don't expect $\tdil{z}{r}$ or $\tdtil{r}{z}$ to be divergent here in a 
general LTB model
with co-ordinate choice \er{RrPNC}.  
Indeed, \er{RhmEq} and 
\er{Rhzzm} are exactly what happens at $\Rh_m$ in the FLRW case.  So although there 
are no divergencies at $\Rh_m$, the numerics break down.  
In Lu and Hellaby \cite{Lu07} this was overcome by doing a series 
expansion in
$\Delta z = z - z_m$, and joining the numerical and series results at some $z$
value $z_a < z_m$---see sections 2.6, 3.3, and appendix B of their paper.  
As pointed out by Hellaby \cite{Hel06}, this phenomenon is not merely a 
cosmological
curiosity.  At this locus, and no other, there is a simple relation between the
diameter distance $d_D = \hat{R}$ and the gravitational mass $M_m$ that is
independent of any inhomogeneity between the observer and sources at this distance:
 \begin{align}
   2 M_m = \Rh_m - \frac{\Lambda \Rh_m^3}{3} ~,   \eqlabel{LamRhmEq}
 \end{align}
 or \er{RhmEq} if $\Lambda = 0$. 
(However, he redshift $z_m$ at which this occurs is model dependent.)
Therefore the maximum in $\Rh$ provides a new characterization of our 
Cosmos---the
cosmic mass.  More importantly for this paper, it provides a cross-check on the
numerical integration, since the $M$ value obtained from the numerical integration
must agree with that deduced from the measured $\Rh_m$ using \er{RhmEq} or 
\er{LamRhmEq}.  For this reason, it plays a prominent role in what follows.

 \subsection{Outline of numerical procedure}

   Since the long-term intent is to use observational data as input, some important 
practical considerations arise, as discussed in Lu and Hellaby \cite{Lu07}.  
The above analysis assumes 
continuous functions and differential equations, whereas real cosmological data consist 
of a large number of discrete measurements of individual sources (e.g.\ galaxies); and 
numerical integration involves discrete steps of finite size.  

   Measurements of the magnitudes (or luminosities) and masses of individual galaxies 
involve considerable uncertainty, so it is advisable to average over a significant sample, 
and this requires putting the data into redshift bins.  
(At present, bins of width $\delta z = 0.001$ are being used.)
While redshift measurements are 
relatively accurate, the observed redshifts include peculiar velocities, so again bin 
averages reduce the uncertainty
 \footnote{
 Actually, the diameter distance is unaffected by peculiar velocity, even if it is calculated 
from the luminosity distance and the redshift using the reciprocity formula \er{dDdL}.  
Therefore, if measurements of diameter distance or luminosity distance become sufficiently 
accurate out to large redshifts, it would be preferable to bin the data by $d_D = \Rh$.  The 
main difficulty would be to distinguish sources just closer and just farther than the maximum in 
$\Rh$.
 }.  The very-low $z$ bins
 \footnote{
 If we could detect all sources, the low-$z$ bins would have very little data, but in fact the 
data are far more complete at low $z$, so the opposite is true.
 }
 are strongly affected by peculiar velocities, and there are no data at
the origin itself, which makes it difficult to set the initial conditions
for the DEs.  Thus, we have to estimate the origin values from the data in
the low-$z$ data bins and a series expansion of the model.  Therefore the
procedure for integrating down the PNC has four main stages: the origin
fitting, the first numerical integration interval, the near maximum
series, and the second numerical integration. 

   Because the DEs involve the first and second derivatives of the
observational quantity $\Rh(z)$, it is evident that noise in this function
could create very large fluctuations in the derivatives.  Thus, in
addition to binning the data, it may be necessary to smooth out
statistical fluctuations by fitting a curve to the binned data. 

   Each of the procedures---binning, smoothing, numerical integration
procedure, etc.---is a potential source of unintended bias and obviously
determines the smallest scale on which inhomogeneity can be detected.  So
it is likely that our thinking about the best way to handle all these
issues will evolve as the project proceeds. 

The next two sections consider the effect of statistical uncertainty in
the data functions $\Rh(z)$ and $4 \pi \mu n(z)$. This affects both the
uncertainty in the output functions $M(z)$, $E(z)$ and $t_B(z)$, and the
stability of the solution procedure. 

\section{Stability of the Differential Equations}

\subsection{Equations for $r$ and $\phi$}

   The co-ordinate distance $r$ is calculated by integrating \er{phiDef}, once 
$\phi = \tdil{r}{z}$ has been determined from \er{dphidz},
\begin{equation}
\frac{\mathrm{d}\phi}{\mathrm{d}z} = \phi_z = \left(\frac{1}{1+z} + 
\frac{\hat{R}_{zz}}{\hat{R}_z} + \frac{4 \pi \mu n \phi}{\hat{R}
\hat{R}_z} \right) \phi 
, \eqlabel{dphidz2}
\end{equation}
where $z$ subscripts denote differentiation with respect to $z$.
The expression for the error in $\phi_z$ due to the error in $\phi$ is
\begin{equation}
\Delta \phi_z = \left( \frac{\phi_z}{\phi} 
+ \frac{4 \pi \mu n \phi}{\hat{R} \hat{R}_z} \right) \Delta \phi ,
\eqlabel{Deltaphi}
\end{equation}
 which holds only for $\Delta \phi \ll \phi$.  
At small $z$, the $4 \pi \mu n \phi/(\hat{R} \hat{R}_z)$ term is 
insignificant relative to the other two terms in $\phi_z$ at $z=0$, 
and $\Rh_z > 0$ before the maximum in $\Rh$, so 
this leads to the integrated values of $\phi$ being stable so long as 
$\phi_z$ is negative and $\phi$ is positive. 
When $\phi_z$ is stable, if $\phi$ is over (or under) estimated, then the 
factor of $\phi$ in $\phi_z$ leads to $\phi_z$ 
being under (or over) estimated so that $\phi$ opposingly decreases (or 
increases) toward the correct value. 
It is possible for $\phi$ to become unstable 
if $\phi_z$ becomes positive;
however, this should only occur if there are strong inhomogeneities, 
and only over a small range of $z$, meaning the 
instability can only grow over a short range before $\phi$ becomes 
stable again and corrects itself.

Beyond the overall influence of the $\phi$ factor in $\phi_z$, there is an additional
$\phi$ in the third term of $\phi_z$ that can lead to instability 
before the maximum in $\hat{R}$.  
As the maximum $\Rh_m$
is approached, the $4 \pi \mu n \phi/\Rh \Rh_z$ term grows faster 
than $\phi_z/\phi$, and after the point where
\begin{equation}
\frac{\hat{R}_{zz}}{\hat{R}_z}
= - \left(
\frac{1}{1+z}+
\frac{8 \pi \mu n \phi}{\hat{R} \hat{R}_z}
\right)   \eqlabel{UnstabPoint}
\end{equation}
the DE becomes increasingly unstable towards $\Rh_m$.  
Near the maximum, the terms in (\ref{dphidz2}) involving $1/\Rh_z$ both diverge, 
but they ought to cancel with ideal data and calculations, 
because $\hat{R}_{zz} = 4 \pi \mu n \phi/\hat{R}$ at $\Rh_m$.
Since the sign of $\hat{R}_z$ must change from positive to negative at $\Rh_m$, 
while all other quantities maintain the same sign, 
then $\phi$ returns to being stable after $\Rh_m$.  

Since the values of $r$ are integrated using the values of $\phi$, the 
stability in 
$r$ depends on the stability in $\phi$.  The main difference is that if 
$\phi$ diverges in one direction and later stabilizes back to 
the proper value, there is a cumulative impact on the integrated 
values of $r$, so the consequence of any instability in $\phi$ is 
potentially worse for $r$.
The integration relies on the use of a series expansion to 
get past the maximum in $\hat{R}$. Provided the integration switches 
over to the series expansion well enough in advance of the maximum, then 
$\phi$ should not diverge significantly from the correct value, and the 
cumulative effect on $r$ should not be significant.

\subsection{Equations for $M$ and $E$}

The differential equation \er{dMdz} with \er{WPNCz} for $M$ is given by
\begin{equation}
\frac{\mathrm{d}M}{\mathrm{d}z} = M_z = 4 \pi \mu n W ,
\eqlabel{dMdz2}
\end{equation}
where
\begin{equation}
W = \sqrt{1+2E} = \frac{\hat{R}_z}{2 
\phi}+\left(1-\frac{2M}{\hat{R}} \right) 
\frac{\phi}{2 \hat{R}_z} ,
\eqlabel{WPNCz2}
\end{equation}
and so the stability equation for $\Delta M \ll M$ is
 \begin{align}
   \Delta M_z = 4 \pi \mu n \left[ \left\{ - \frac{\Rh_z}{2 \phi^2}
   + \left( 1 - \frac{2 M}{\Rh} \right) \frac{1}{2 \Rh_z} \right\} \Delta \phi
   - \frac{\phi}{\Rh \Rh_z} \Delta M \right] ~.
 \end{align}
Since $W$ contains a term that goes as $-M \phi/(\hat{R} \hat{R}_z)$, if 
$M$ is over (or under) estimated, then $W$, $E$, $M_z$, and $M$ are 
opposingly under (or over) estimated when $\hat{R}_z$ is positive, making the 
determination of $M$ from $W$ stable and the determination of $W$ (and $E$) 
from $M$ stable. 
Barring the 
presence of strong inhomogeneities, $\hat{R}_z$ should remain positive out to 
the maximum in $\hat{R}$, so $M$ and $W$ should remain stable with each 
other up to the maximum. If there are strong inhomogeneities, these 
should only make 
$\hat{R}_z$ go negative over short regions of the integration, meaning 
the instability can only grow briefly before $M$ and $W$ become stable 
and return to their proper values.
The values of $W$, $E$, $M_z$, and $M$ also depend on the integrated 
values of $\phi$, so if $\phi$ becomes unstable, as it 
may do before the maximum, then $W$ and $M$ can potentially become 
unstable due to $\phi$, but as discussed previously, the series expansion 
of $\phi$ around the maximum should bridge this instability in $\phi$.

After the maximum in $\hat{R}$, $\hat{R}_z$ should remain negative (except in 
the presence of strong inhomogeneities), which means over (or under) 
estimates in $M$ lead to over (or under) estimates in $W$ and $E$, 
and over (or under) estimates in $W$ lead to over (or 
under) estimates in $M_z$ and $M$. Thus, $M$ and $W$ feed back on each 
other and lead to instability. 
This suggests an alternative method is needed to determine $M$ and $E$ 
after the maximum in $\hat{R}$ is reached. Assuming the data do not 
extend significantly beyond the maximum, one method is to 
create an extended series expansion in $W$ based on the $4 \pi \mu n$ 
and $\hat{R}$ data from the maximum to the outer boundary, and then 
integrate values of $M$ based on the $W$ series expansion so that $M$ and 
$W$ do not feed back on each other.
If the data extend significantly beyond the maximum, and no alternative 
method is found, we must expect very large uncertainties in $W$ and $M$.

\section{Numerical Handling of Statistical Errors}

\subsection{Simulating errors with Gaussian deviates}

In order to make sure the code can handle real data, statistical errors 
must be added to the distance determinations and galaxy number counts for 
the test data. 

The actual number of sources $N$ per bin is equal to $4 \pi n \delta z$ 
(where $\delta z$ is the redshift interval of the bin), since $n$ is the number
of sources per steradian and per unit redshift interval.  From the test 
universes, the number of sources $N$ is calculated as 
\begin{equation}
N = \frac{4 \pi \mu n \delta z}{\mu} ,
\end{equation}
 where $\mu$ is the assumed mass of a typical galaxy---$10^{11}~M_\odot$.
The program then calculates the expected errors in the $\hat{R}$ and $4 \pi 
\mu n$ values by calculating the $1$-$\sigma$ errors and using the 
elimination method to add Gaussian deviates (e.g.\ \cite{Pre02}) to the 
data in a $\pm 5 \sigma$ range. 

The random error in a single $\hat{R}$ measurement is assumed to be of order
10\% so that the random error $\Delta \hat{R}$ in the $\hat{R}$ value 
for a bin of $N$ galaxy sources is
\begin{equation}
\Delta \hat{R} = \frac{0.1 \hat{R}}{\sqrt{N}} .
\end{equation}
 (Although current luminosity-distance measurements are mostly not this good, 
it is expected the accuracy will improve considerably in the coming years.)  

The random error in the $4 \pi \mu n$ measurements is
\begin{equation}
\Delta (4 \pi \mu n) 
= \frac{1}{\delta z} \sqrt{(\Delta \mu N)^2+(\Delta N \mu)^2}
= \frac{1}{\delta z} \sqrt{(\Delta m \Delta N)^2+(\Delta N \mu)^2}
, 
\end{equation}
where $\Delta N = \sqrt{N}$, and $\Delta \mu = \Delta m/ \Delta N$ relates
the error in the mean mass per source $\mu$ and the error in a single mass
measurement $m$.  The uncertainty in $N$ is calculated using the Poisson
error, since if the galaxies occur randomly, then the uncertainty in the
number that should be observed goes as $\sqrt{N}$. (In reality galaxies
are clustered, but since we are binning over thin spherical shells, 
clusters will be averaged out at large radii.) Assuming $\Delta m$ is
less than (or of the same order as) $\mu$, then the random error in the
$4 \pi \mu n$ measurements is approximately
\begin{equation}
\Delta (4 \pi \mu n) 
= \frac{\mu \sqrt{N}}{\delta z} 
= \frac{4 \pi \mu n}{\sqrt{N}} .
\end{equation}

\subsection{Calculating derivatives of $\hat{R}$}

With errors in the $\hat{R}$ values, it is no longer possible to use the
original method of Lu and Hellaby \cite{Lu07} of calculating the first and 
second derivatives, 
which were obtained from the first and second differences in $\hat{R}$ between 
adjacent redshift bins.  Due to errors in the data, the
noise in the second derivative ends up being of order 10000\%.
This causes the integrated values of $\phi$ to vary rapidly up and down and
quickly diverge shortly after the instability in $\phi$ develops (before
the maximum in $\hat{R}$). 

Thus, in order to reduce the noise in the derivatives, it is necessary to
calculate least-squares fits to the $\Rh$ data.  
Unfortunately, since the functional form of $\Rh(z)$ cannot be assumed, simply 
fitting a single polynomial is unlikely to yield a good fit over the whole 
range of data. 
Rather, the values of the first and second derivatives that best reflect the 
local data are obtained by fitting a polynomial to the $2 k + 1$ redshift bins 
centered on each $z_i$, covering the range $z_i \pm k \delta z$.
Since variations in the second derivative should
exist, especially if strong inhomogeneities are present, it is necessary
to use at least a cubic polynomial, so that the second derivative is
allowed to vary over the range of the fit. Near the
origin and outer boundary, it is not possible to have a range of data
that is centered exactly on the point in interest, making it even more
important to have at least a quartic fit so that the unequal upper
and lower ranges of data do not bias the calculation of the second
derivative. 

When there are strong variations in $\hat{R}_{zz}$, it is necessary to either
reduce the range of data used in the fit or increase the order of the fit
so that they do not get smoothed out. In either case
this leads to more noise in the fit. The ideal data range and polynomial
order varies depending on the test universe. If there are no strong
variations it is better to use a lower order polynomial (such as a quartic) 
and a large range of data (up to 300 redshift bins centered on the 
current $z$ value)
to reduce the noise, but when there are strong variations, it is better to
increase the order or reduce the range to allow some noise, in the
interest of fitting the actual variation.  It is obvious if real
variations are present, since these are noticeable even in the 
lower-order fits where the noise is
negligable. Exactly how much the range should be reduced or the order of
the polynomial should be increased can be ascertained by changing the fit
until the real variations cease to become stronger or when
the noise starts to become a significant fraction of the magnitude of the
real variation.  In practice, for the test model with strongest inhomogeneity, 
the range could be as small as 30 $\delta z$ bins and the fit as high as 7th order.

More precisely, the best polynomial fit to the data can be ascertained by 
comparing the ratio of $\chi^2$ with the number of degrees of freedom (NDF), 
where $\chi^2$ is the sum of the squares of the $\Rh$ residuals divided by 
the 1-$\sigma$ error of each $\Rh$ value, and the NDF
is the number of data points minus the number of polynomial coefficients 
being fit. 
The expected value of $\chi^2$ is the NDF, so 
the ratio of the two that is closest to 1 yields the best fit to the data.
A ratio greater than 1 is not adequately fitting the data, while a ratio 
smaller than 1 is fitting the data too well, which means it is likely 
fitting the noise and is not the most realistic fit to the data.

It should be noted that the least-squares fitting of $\hat{R}$ versus $z$
minimizes the error in the $\hat{R}$ values, so it assumes no error in
$z$.  Any actual errors in $z$ would bias the fit.
While the observational errors in $z$ should be small, the peculiar
velocity components of the observed redshifts may lead to errors. 

Another consequence of this smoothing is that random fluctuations 
in the $\Rh$ data create quasi-periodic variations 
in the smoothed versions of $\Rh$, $\Rh_z$, and $\Rh_{zz}$, 
which have a period determined by the $z$ range chosen for 
the fit.  These manifest themselves in $\phi$ as oscillations 
with an amplitude that diverges as the maximum $\Rh_m$ is approached.

\subsection{Using an origin LTB series expansion}

The integration of the differential equations must proceed from an initial value,
which in the original code of Lu and Hellaby \cite{Lu07} involved using an
FLRW series expansion at the origin to determine the derivatives in the
first two redshift bins.
Since a regular LTB model is FLRW-like at the origin, this is not a 
bad first approximation.
However, the fractional errors in $\hat{R}$ and $4 \pi \mu n$ are
largest near the origin, due to the smaller number counts that occur in a
bin of given thickness $\delta z$, and also, the $\hat{R}$ derivatives are
more susceptible to error near the origin. The determination of the bang
time is more difficult near the origin, because this depends on
$2E\hat{R}/M$, while $E$, $\hat{R}$, and $M$ are all approaching
zero and highly susceptible to numerical errors. Thus, it is preferable to use 
the series expansion for more than just the first two bins, which 
necessitates the use of an LTB series so that the region around the origin 
is not constrained to be FLRW.  
Approximating $\Rh$ and $4 \pi \mu n$ near the origin by 5th order power 
series in $z$, the series coefficients are found by a least-squares fit to 
the data in the first 50 bins (which extends to $z=0.05$---a distance 
scale at which there should be adequate data to begin integrating from bin 
to bin without large numerical errors).
 The functions $\phi$, $M$, and $E$ are also expressed as power series, and their 
coefficients are derived from the DEs.  The values of the functions at the end of 
the origin series then provide the initial conditions for the numerical integration.

Series expansions of the $\hat{R}$ and $4 \pi \mu n$ data near the 
origin in the form
\begin{align}
\hat{R}(z) = R_1 z + R_2 z^2 + R_3 z^3 + R_4 z^4 + R_5 z^5
\end{align}
and
\begin{align}
4 \pi \mu n (z) = K_2 z^2 + K_3 z^3 + K_4 z^4 + K_5 z^5
\end{align}
are used to calculate the series expansions of $r$, $\phi$, $M$,
and $2E$ from \er{dphidz}, \er{phiDef}, \er{dMdz}, and \er{WPNCz}, as
\begin{align}
\begin{array}{c}
r(z) = R_1 z+\left(\frac{R_1}{2}+R_2 \right) z^2 + 
       \left(\frac{2 R_2}{3}+R_3+\frac{K_2}{6} \right) z^3 
       +\left(\frac{3 R_3}{4}+\frac{5 K_2}{24}+R_4+\frac{K_3}{12}
       +\frac{K_2 R_2}{6 R_1} \right) z^4           \\
+\left(
R_5
+\frac{K_4}{20}
+\frac{7 K_3}{60}
+\frac{K_2}{15}
+\frac{4 R_4}{5}
+\frac{ 
\left(\frac{K_3}{12}+\frac{13 K_2}{60} \right) R_2
+\frac{K_2 R_3}{4} 
+ \frac{{K_2}^2}{20}
}{R_1}
-\frac{K_2 {R_2}^2}{12 {R_1}^2} 
\right) z^5 ,
\end{array}
\end{align}

\begin{equation}
\begin{array}{c}
\phi(z) = R_1 + 2 \left(\frac{R_1}{2}+R_2 \right) z + 3
       \left(\frac{2 R_2}{3}+R_3+\frac{K_2}{6} \right) z^2
       +4 \left(\frac{3 R_3}{4}+\frac{5 K_2}{24}+R_4+\frac{K_3}{12}
       +\frac{K_2 R_2}{6 R_1} \right) z^3           \\
+ 5 \left(
R_5
+\frac{K_4}{20}
+\frac{7 K_3}{60}
+\frac{K_2}{15}
+\frac{4 R_4}{5}
+\frac{ 
\left(\frac{K_3}{12}+\frac{13 K_2}{60} \right) R_2
+\frac{K_2 R_3}{4} 
+ \frac{{K_2}^2}{20}
}{R_1}
-\frac{K_2 {R_2}^2}{12 {R_1}^2} 
\right) z^4 ,
\end{array}
\end{equation}
\begin{equation}
\begin{array}{c}
M(z) = \frac{K_2}{3} z^3 + \frac{K_3}{4} z^4
+\left(
\frac{-{K_2}^2}{15 R_1}
+\frac{K_2}{10}
+\frac{K_4}{5}
\right) z^5 \\
+\left(
\left(\frac{1}{36 R_1}+\frac{R_2}{18 {R_1}^2} \right) {K_2}^2
+\left(\frac{-1}{12}-\frac{7 K_3}{72 R_1} \right) K_2
+\frac{K_3}{12}
+\frac{K_5}{6} 
\right) z^6 
,
\end{array}
\end{equation}
and
\begin{equation}
\begin{array}{c}
E(z) = \left( \frac{1}{2}-\frac{K_2}{3 R_1} \right) z^2
+\left(
\left(
\frac{1}{6 R_1}+\frac{R_2}{3 {R_1}^2}
\right) K_2-\frac{1}{2}-\frac{K_3}{4 R_1}
\right) z^3 \\
+\left(
\frac{29 {K_2}^2}{360 {R_1}^2}
+\left(
-\frac{11}{60 R_1}+\frac{R_3}{3 {R_1}^2}
-\frac{{R_2}^2}{3 {R_1}^3}
\right) K_2
+\left(
\frac{1}{12 R_1}+\frac{R_2}{4 {R_1}^2}
\right) K_3+\frac{5}{8}-\frac{K_4}{5 R_1}
\right) z^4 \\
+
\left(
\left(
-\frac{1}{360 {R_1}^2}-\frac{11 R_2}{90 {R_1}^3}
\right) K_2^2 
\right. 
+\left(
\frac{K_3}{9 {R_1}^2}+\frac{19}{60 R_1}+\frac{\frac{11 R_2}{60}+\frac{R_4}{3}+\frac{R_3}{12}}{{R_1}^2}
+\frac{-\frac{2 R_2 R_3}{3}-\frac{{R_2}^2}{12}}{{R_1}^3}
+\frac{{R_2}^3}{3 {R_1}^4}
\right) K_2 \\
\left.
+\left(
-\frac{1}{8 R_1}+\frac{R_3}{4 {R_1}^2}-\frac{{R_2}^2}{4 {R_1}^3}
\right) K_3
+\left(
\frac{1}{20 R_1}+\frac{R_2}{5 {R_1}^2}
\right) K_4
-\frac{3}{4}-\frac{K_5}{6 R_1}
\right) z^5
.
\end{array}
\end{equation}

In Figure \ref{tBErrors}, the relative error in the 
bang time near the origin is 
shown for both the previous method and the new method.  The large numerical 
error that occurs with the previous method is greatly reduced by using the LTB 
series expansion.

\begin{figure}
\begin{center}
\resizebox{88mm}{!}
{\includegraphics{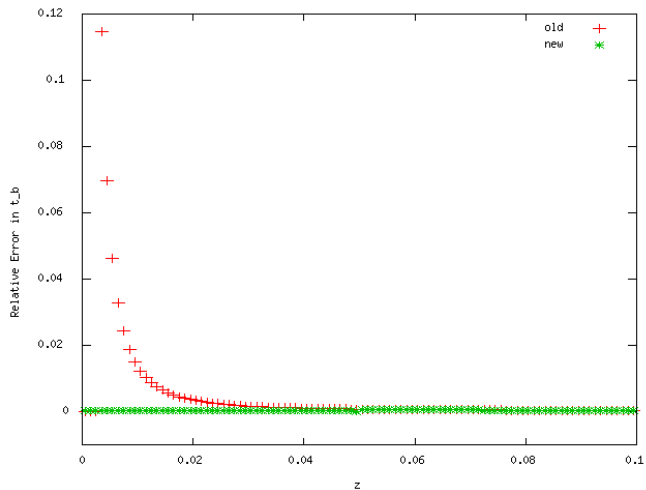}}
\end{center}
\caption{
 \figlabel{tBErrors}
 Comparison of the relative errors in the bang time $t_B$ versus $z$ using the 
old (crosses) and new (asterisks) methods.
The new method fits an LTB series expansion to the first 50 points, 
before starting the numerical integration, thus allowing the solution to be 
inhomogeneous while avoiding the numerical difficulties near the origin.
This plot is for an inhomogeneous universe with origin 
Hubble and deceleration parameters $h_0=0.72$ and 
â$q_0=0.6$, and with $\Omega_{\Lambda}=0$.} 
\end{figure}

\subsection{Bridging the maximum in the areal distance}

 Since, as noted in Section \ref{ApHor}, $\hat{R}_z$ goes to zero at $\Rh_m$ (the 
maximum in $\hat{R}$), and the DEs for $\phi$ \er{dphidz} and $M$ \er{dMdz} 
and \er{WPNCz} both depend on $1/\Rh_z$, it is necessary to use a series expansion
in the neighborhood of the maximum.  Also, the $\phi$ DE becomes unstable before 
the maximum, so this series expansion is necessary over a range of more than just a 
few points.  

Appendix \ref{NearMaxSeries} gives the series solutions for $\phi$, $M$, and 
$W$ in powers of $\Delta z = z_m - z$.  
The coefficients in the $\phi$ series are 
fully determined by the DEs, but the $M$ and $W$ coefficients are interrelated and 
contain one undetermined coefficient, the linear term in the $M$ series, or 
equivalently, the constant term in the $W$ series.  In fact every coefficient of 
the $M$ and $W$ series depends linearly on this factor, 
so its value is easily fixed by matching 
these two series to their numerical values at some cross-over point $z = z_a < z_m$, 
where the series and numerical $\phi$ curves are in closest agreement, as was done 
by Lu and Hellaby \cite{Lu07}.  A second matching is done at $z = z_J > 
z_m$, where 
the near-maximum series ends and the numerical integration is re-started.

When statistical errors are present however, the oscillations in $\phi$ that grow 
as $\Rh_m$ is approached make the matching of the numerical and series $\phi$ values 
problematic, since they will tend to intersect at multiple points. Thus, 
the code has been modified such that the two 
$\phi$ curves are matched by calculating the absolute value of the differences 
between the curves over a range of the data (the same range used in the 
least-squares fit of $\hat{R}$) and matching them at the $z$ value where the total 
difference is minimized. 
This generally leads to a matching near where
the $\phi$ instability sets in, thus excising the region of diverging $\phi$ error.

In the original code of Lu and Hellaby \cite{Lu07}, $M$ was matched at the 
start of the series, and $W$ was matched at the end of the series. 
While $M$ and $W$ are stable with each other before $\Rh_m$, the 
fluctuations in $\phi$ influence $W$, which opposingly influences $M$, so 
typically there is a small error in $M$ at the first matching point.  
This $M$ matching error biases the slope 
of the $M$ series, and in turn the constant term of the $W$ series. Thus, 
there is generally a small jump in $W$ at the first matching point and a 
small jump in $M$ at the second matching point, which in turn leads to a 
rapid adjustment in $W$ once the numerical integration re-starts.  

However, it turns out that it is possible to find values of $M$ and $W$ 
that are consistent with each other at the first matching point so that 
the small error in $M$ does not lead to overcorrections. 
The mean of the integrated value of $M$ and the value of $M$ that would 
result from matching the integrated value of $W$ is used to determine the 
series-connection value of $M$.  Similarly, 
the mean of the integrated value of $W$ and the value of $W$ that would 
result from matching the integrated value of $M$ determines the
series-connection value of $W$.
These new series-connection values of $M$ and $W$ are consistent with 
each other in that calculating the $M$-series matching using the mean 
$W$ value for the matching leads exactly to the mean $M$ value, and vice 
versa for the mean $W$ value.  Also, these values are stable in that
calculating the matching with ``numerical" values above (or below) the mean results in 
series-connection values closer to the mean values.
This can be explained by the fact that $M$ and $W$ are stable with each 
other before $\Rh_m$, so there should be stable series expansion values 
of $M$ and $W$ that are consistent with each other before $\Rh_m$.

Plots of the relative error in $M$ and $W$ in the region containing the 
matching before $\Rh_m$ appear in Figure 
\ref{WMErrors}.  Using this new matching method
reduces the jump in $W$ at the first matching so that it does 
not overcorrect, and it leads to more accurate values of $W$
throughout the series expansion. 
It also leads to more accurate values of $M$ in the near-maximum series, 
and much better values of $M$ at the second matching. 
This is an important improvement over the original code, especially when 
there are errors in the data that can lead to larger jumps in $M$ and $W$. 

\begin{figure*}
\resizebox{88mm}{!}
{\includegraphics{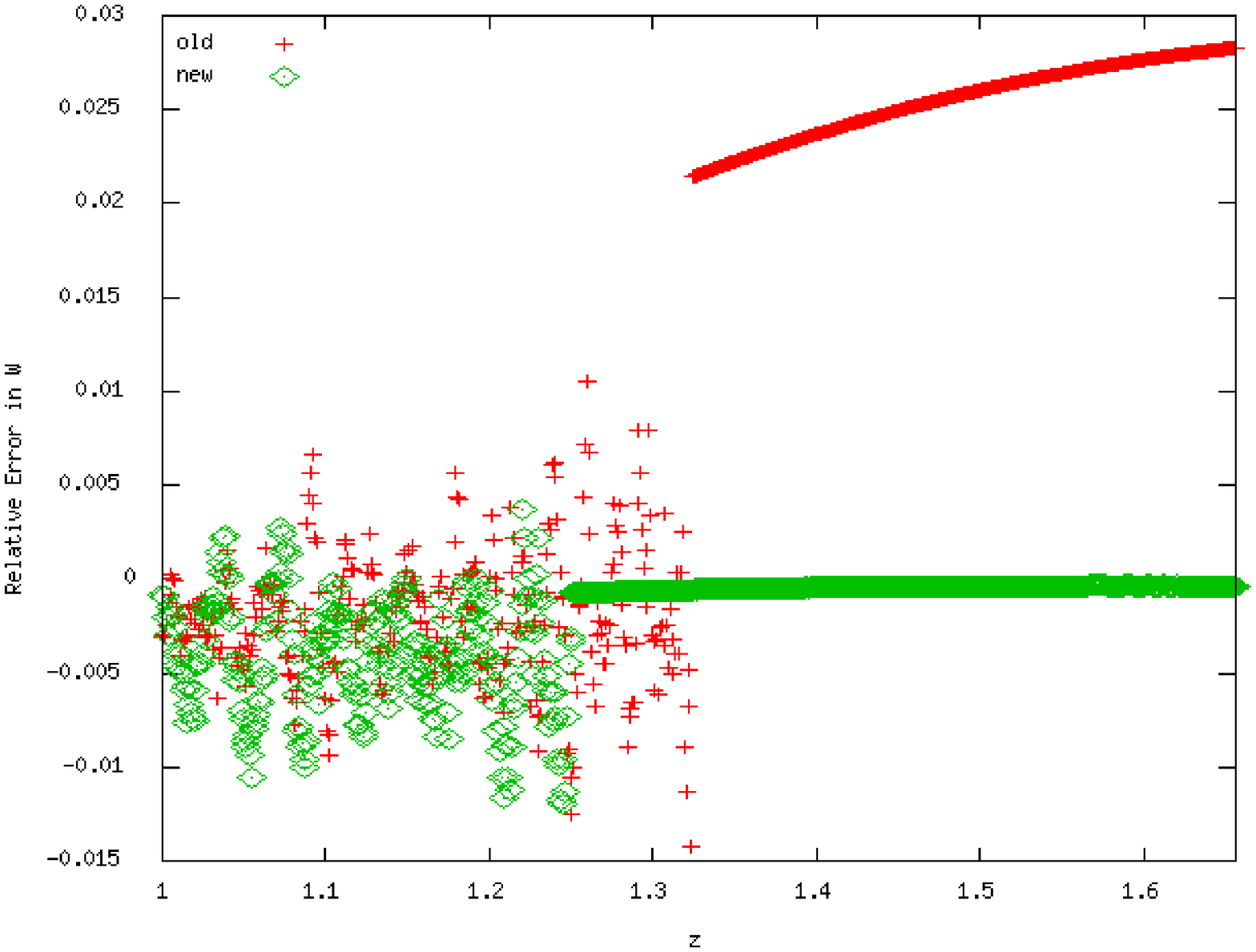}}
\resizebox{88mm}{!}
{\includegraphics{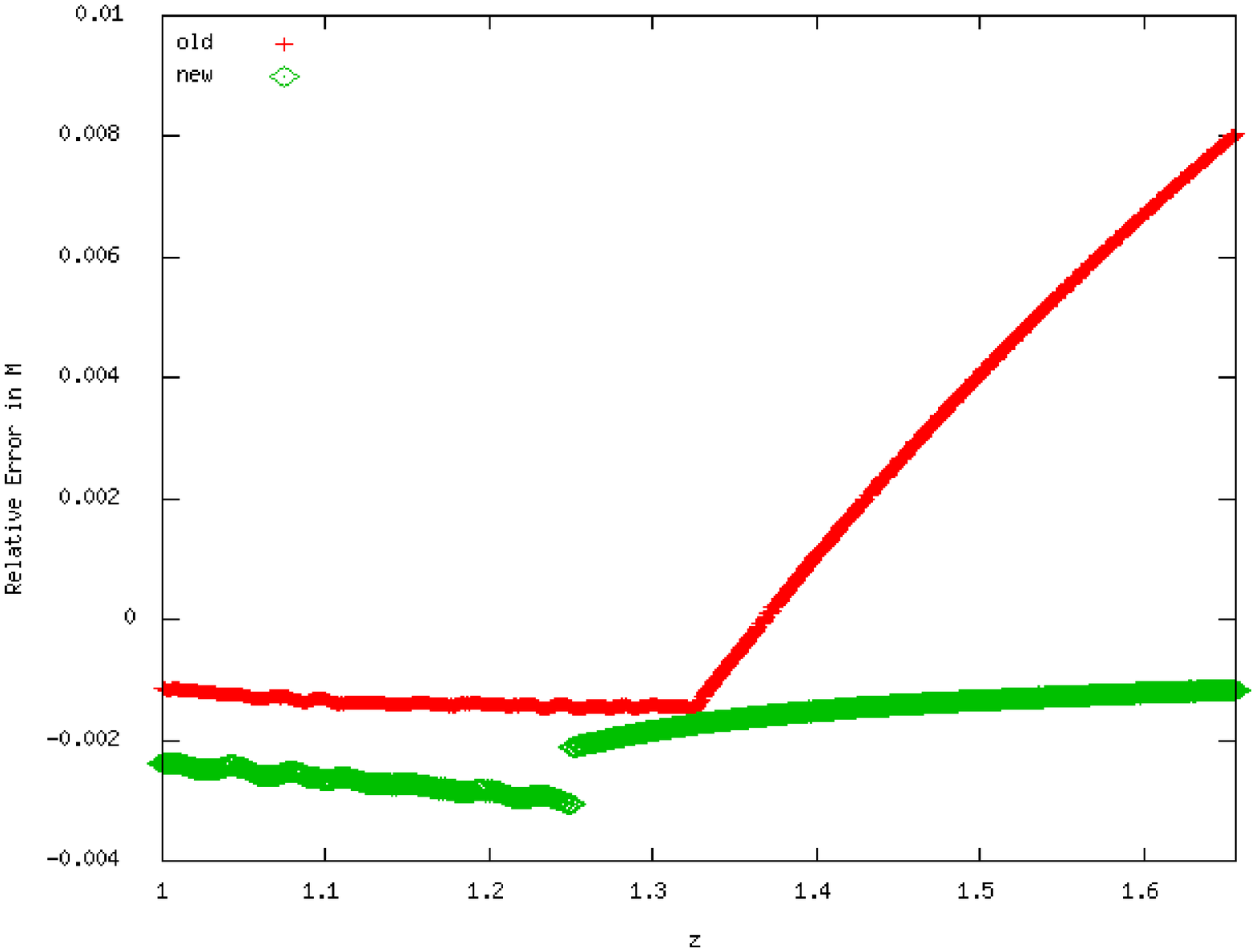}}
\caption{
 \figlabel{WMErrors}
 Comparison of the relative errors in $W$ (left) and $M$ (right) versus $z$ 
in the region before $\Rh_m$, using the old (small
crosses) and new (large diamonds) methods for matching 
from the numerical values to the near-maximum series expansion values.
These plots are for an FLRW universe with $h_0=0.72$, $q_0=0.22$, and 
$\Omega_{\Lambda}=0$, with statistical errors added to the data to 
simulate observational errors.
Simply matching $M$ can lead to a large jump in $W$ and a divergence in 
$M$, whereas the new method leads to small jumps in both $M$ and $W$ 
that always bring them toward the correct values.}
\end{figure*}

\subsection{Propagation of uncertainties}

To estimate the uncertainties, the observational errors need to be taken
into account in how they propagate in the integrations.  
Where a series fit to the data is performed, the estimated errors come from 
the covariance matrix of the fit.  The uncertainties in the numerical integration of 
 $r_z$, $\phi_z$, $W$, and $M_z$ are
\begin{equation}
\Delta r_z = \Delta \phi,
\end{equation}
\begin{equation}
\begin{array}{c}
\eqlabel{52}
\Delta \phi_z = 
\left(
\left[
\left(
\frac{1}{1+z}+\frac{\hat{R}_{zz}}{\hat{R}_z}
+\frac{8 \pi \mu n \phi}{\hat{R} \hat{R}_z}
\right) \Delta \phi
\right]^2
+
\left(
\frac{\phi}{\hat{R}_z} \Delta \hat{R}_{zz}
\right)^2
+
\left[
\left(
\frac{\hat{R}_{zz} \phi}{ {\hat{R}_z}^2 }+
\frac{ 4 \pi \mu n {\phi}^2 }{ \hat{R} {\hat{R}_z}^2 }
\right) \Delta \hat{R}_z
\right]^2 
\right. \\
+
\left.
\left(
\frac{4 \pi \mu n {\phi}^2}{\hat{R}^2 \hat{R}_z} \Delta \hat{R}
\right)^2 
+\left(
\frac{{\phi}^2}{\hat{R} \hat{R}_z} \Delta (4 \pi \mu n) 
\right)^2
+\left(
\frac{1}{(1+z)^2} \Delta z 
\right)^2 
\right)^{1/2}
, 
\end{array}
\end{equation}
\begin{equation}
\begin{array}{c}
\Delta W = 
\left(
\left[
\left(
\frac{\hat{R}_z}{2 \phi^2}-\frac{1-2M/\hat{R}}{2 \hat{R}_z}
\right) \Delta \phi
\right]^2 +
\left[
\left(
\frac{1}{2 \phi} - \frac{(1-2M/\hat{R})\phi}{2{\hat{R}_z}^2}
\right)
\Delta \hat{R}_z
\right]^2
\right. 
\left. +
\left(
\frac{M \phi}{\hat{R} {\hat{R}_z}^2} \Delta \hat{R}
\right)^2 +
\left(
\frac{\phi}{\hat{R} \hat{R}_z} \Delta M
\right)^2
\right)^{1/2}
,
\end{array}
\end{equation}
and
\begin{equation}
\Delta M_z = 
\left(
\left[
\left(
W \Delta (4 \pi \mu n) 
\right)^2
+ 
\left(
4 \pi \mu n \Delta W
\right)^2
\right]
\right)^{1/2}
,
\end{equation}
assuming that the errors are uncorrelated with each other so that they 
are added in quadrature.

The $\phi_z$ values are used to calculate each $\phi$ value from the previous
$\phi$ value, but how the error in $\phi$ accumulates depends on the
stability in $\phi$. It would be expected that when $\phi$ is stable this
error would not grow from step to step, while when $\phi$ is unstable the
errors from all the unstable steps would need to be added together directly.
In practice, calculating the $\phi$ uncertainty is more complicated, due to
the smoothing of the $\Rh$ data.  The errors in the smoothed $\hat{R}_{zz}$
values tend to fluctuate up and down with a period $\zeta = 2k \, \delta z$
equal to the range of data that is used to perform the least-squares fit.
These fluctuations in the $\hat{R}_{zz}$ errors induce fluctuations in the
$\phi$ values, so even when $\phi$ is stable it tends to oscillate with the
same period as the range of the least-squares fit.  However, over a range
$\geq \zeta$, the errors due to the fluctuations tend to cancel, and so
should not be added continuously. 

The actual calculation of the $\phi$ uncertainty proceeds by taking
the errors of all the previous steps that may potentially be continuously 
fluctuating in one direction (starting either from the initial point of 
integration or at a point that is previous to the point of interest $z_i$ by half 
the range of the $\hat{R}$ fit, $\zeta/2$) and adding these in quadrature (since the 
errors at these steps are unlikely to all be of the same sign) to get the error
due to the fluctuations $\Delta \phi_f$:
 \begin{align}
   \Delta \phi_f = \sum_{j=i-k}^{i} \Delta \phi_z[j] \, \delta z ~
,
   \eqlabel{Deltaphi-f}
 \end{align}
and combining this error with the error $\Delta \phi_s$ from the end of the series leading up to the 
numerical integration by adding them in quadrature:
 \begin{align}
   \Delta \phi = \sqrt{ (\Delta \phi_s)^2  + (\Delta \phi_f)^2}\
.
   \eqlabel{Deltaphi-q}
 \end{align}
The error $\Delta \phi_f$ is
due to all the uncertainties in the data (\ref{52}) 
over half the smoothing range (or less if the start of the numerics occurs within that range).
The junction of the numerical region to the near-maximum series expansion generally 
occurs just where $\phi$ becomes unstable and $\Delta \phi$ starts to diverge, so it is essentially
unnecessary to ever start adding the full $\Delta \phi$ errors directly instead of in quadrature.

It is assumed that the errors coming out of the origin series are zero, 
since in practice the actual errors are always miniscule compared with the 
estimated error beyond the first few points of the series.  
Beyond the maximum, in the second numerical integration region, the 
error due to the near-maximum series is not negligible, 
but this is slowly corrected by the natural stability of the $\phi$ DE, so that it decays 
from step to step as
 \begin{align}
   \Delta \phi_s = \Delta \phi [i_0] - \sum_{j=i_0}^{i} \Delta \phi_z^* [j] \, \delta z
,
   \eqlabel{Deltaphi-s}
 \end{align}
where $\Delta \phi [i_0]$ is the error at the end of the near-maximum series, 
and $\Delta \phi_z^*$ is calculated from (\ref{Deltaphi}), reflecting how the error
in $\phi$ acts to stabilize itself.  
Figure \ref{phierr} contains a plot showing how the 
estimated error 
compares with the actual error in the propagation of the $\phi$ uncertainties.

\begin{figure}
\begin{center}
\resizebox{88mm}{!}
{\includegraphics{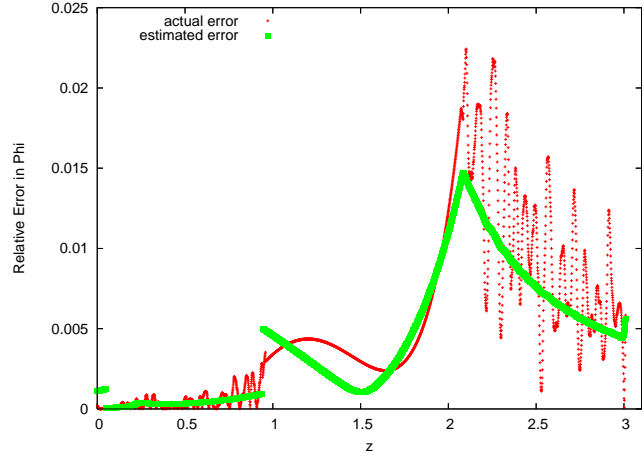}}
\end{center}
\caption{
 \figlabel{phierr}
Relative $\phi$ error versus $z$, with the estimated error (large points)
plotted alongside the actual error (small points) for comparison. Gaussian
deviates are added to the data, and the estimated uncertainty calculated,
along with the actual error from the ideal data. It can be seen that the
error in $\phi$ grows before the maximum as it becomes increasingly unstable,
and then decreases after the maximum as it becomes stable again. This plot is
for an inhomogeneous universe with origin parameters $h_0=0.72$, $q_0=0.6$,
and $\Omega_{\Lambda}=0$.}
\end{figure}

The integrated $r$ values come directly from the $\phi$ values, but the
uncertainties have to take into account the uncertainties in all the $\phi$
values. Even when $\phi$ is stable and self-corrects when an error emerges,
the integrated values of $r$ all continue to be off by the amount gained or
lost while the values of $\phi$ were in error. When $\phi$ fluctuates above
and below the correct value, then the errors in the $r$ values grow like a
random walk. Thus, the error in $r$ is calculated by adding all the previous
errors due to $\phi$ in quadrature to account for how far it should have
wandered due to the random high and low values of $\phi$. 

The integrated values of $M$ are stable before the maximum in $\hat{R}$ and
then become unstable after the maximum due to the feedback with $W$.  Thus,
before the maximum, it is assumed that there is no $M$ error in the
calculation of the $W$ error, and no $W$ error in the calculation of the $M$
error, so that these errors do not feed back on each other and depend solely
on the $4 \pi \mu n$, $\phi$, $\hat{R}$, and $\hat{R}_{z}$ errors. After the
maximum, the $W$ errors are included in the calculation of the $M_z$ errors,
and if the $W$ values are being determined from the integrated values of $M$
(rather than an extended series expansion for $W$), then the $M$ errors need
to be included in the calculation of the $W$ errors as well. 

\section{Dealing with Systematic Errors}

It is very likely that there will be systematic errors in the observations as
well as the evolution functions.  We here investigate the effect of such
errors, and show that the data functions $\Rh(z)$ and $4 \pi \mu(z) n(z)$
must satisfy a consistency condition.  We demonstrate how a discrepancy in
this condition can be corrected in order to make the data and the numerical
result self-consistent.  However, it is not possible to determine from this
process how much error is attributable to $\Rh$ and how much to $4 \pi \mu
n$, or to discern whether the systematic errors vary with $z$. 

\subsection{Systematic error in the distance scale}

We first consider a constant percentage error in $\hat{R}$ only.  In this
case, the origin values of $\phi$ start correspondingly high or low, since
$\phi$ goes as $\hat{R}/z$ near the origin.  Then by \er{dphidz2} the
numerical integration of $\phi$ continues to be off since $\phi_z$ depends on
a $\hat{R}_{zz} \phi/\hat{R}_z$ term.  When the $4 \pi \mu n \phi/(\hat{R}_z
\hat{R})$ term in $\phi_z$ becomes significant, $\phi$ starts to correct
itself since the errors in $1/(\Rh_z \Rh)$ bias the integration of $\phi$ in
the opposite sense, and the integration may even overcorrect if it proceeds
long enough. 

By \er{NMSphi0}-\er{NMSphi4} the erroneous $\Rh$ values produce
correspondingly high or low values in the near-maximum series for $\phi$, but
because of the correction that takes place before the maximum, there is
generally a discrepancy between the numerical and series $\phi$ values so a
close matching is not possible.  This discrepancy is a clear indicator of an
overall systematic error.  Beyond the near-maximum series, the numerical
integration is stable and starts to correct the value of $\phi$, leading to a
discontinuity in the slope of the $\phi$ curve, providing a secondary
indication of a systematic error.  Since the integrated values of $r$ depend
on the $\phi$ values, then if $\phi$ is too high (or low) over most of the
run, then $r$ tends to be correspondingly high (or low) over the whole run by
approximately the percentage of the error in the $\hat{R}$ values, but it is
not obvious from the $r(z)$ curve that a systematic error is present. 

The presence of a jump in $\phi$ at the start of the near-maximum series not
only flags a systematic error, but also provides the means of correcting it.
Thus, simply by systematically increasing or decreasing all of the $\hat{R}$
values until the jump in $\phi$ is minimized, the best correction for
$\hat{R}$ can be found.  The adjustment of the $\Rh$ data necessitates a
repeat of the entire integration from the origin to the maximum series
matching point. The effects of a systematic error in $\hat{R}$ on $\phi$ and
$r$ can be seen in Figure \ref{SystRhErr}. 

A systematic error in $\hat{R}$ is also apparent in $W$. Initially, the
values of $W$ and $M$ are unaffected by the $\hat{R}$ values.  The values of
$\phi$, $\hat{R}$, and $\hat{R}_z$ are generally all correspondingly high or
low, so when the $-M \phi/(\hat{R} \hat{R}_z)$ term in \er{WPNCz2} starts to
become significant, $W$ becomes correspondingly too high or too low (and $M$
in turn also becomes slightly too high or too low).  However, when matching
the numerical integration to the near-maximum series, the change in the value
of $\Rh_m/2$ is greater than the percentage change in $M$, which influences
the slope of the $M$ series and the constant term of the $W$ series such that
$M$ and $W$ end up being correspondingly high or low in the ensuing numerical
integration.  Since the calculation of $t_B$ depends on $E$ and $M$, the
jumps in $W$ and $M$ also typically lead to jumps in $t_B$. The effects of a
systematic error in $\hat{R}$ on $W$, $M$, and $t_B$ can also be seen in
Figure \ref{SystRhErr}. 

\begin{figure*}
\resizebox{88mm}{!}
{\includegraphics{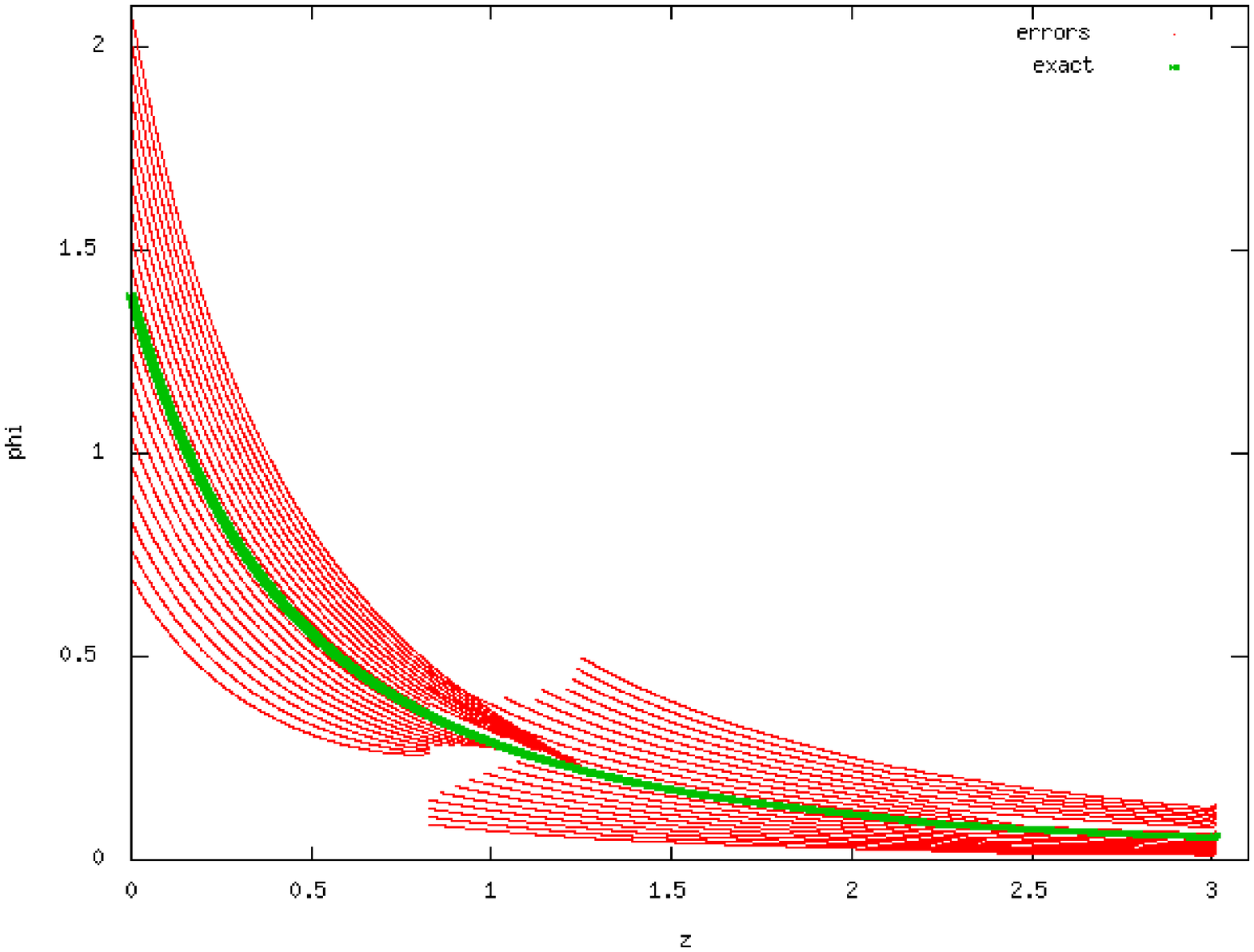}}
\resizebox{88mm}{!}
{\includegraphics{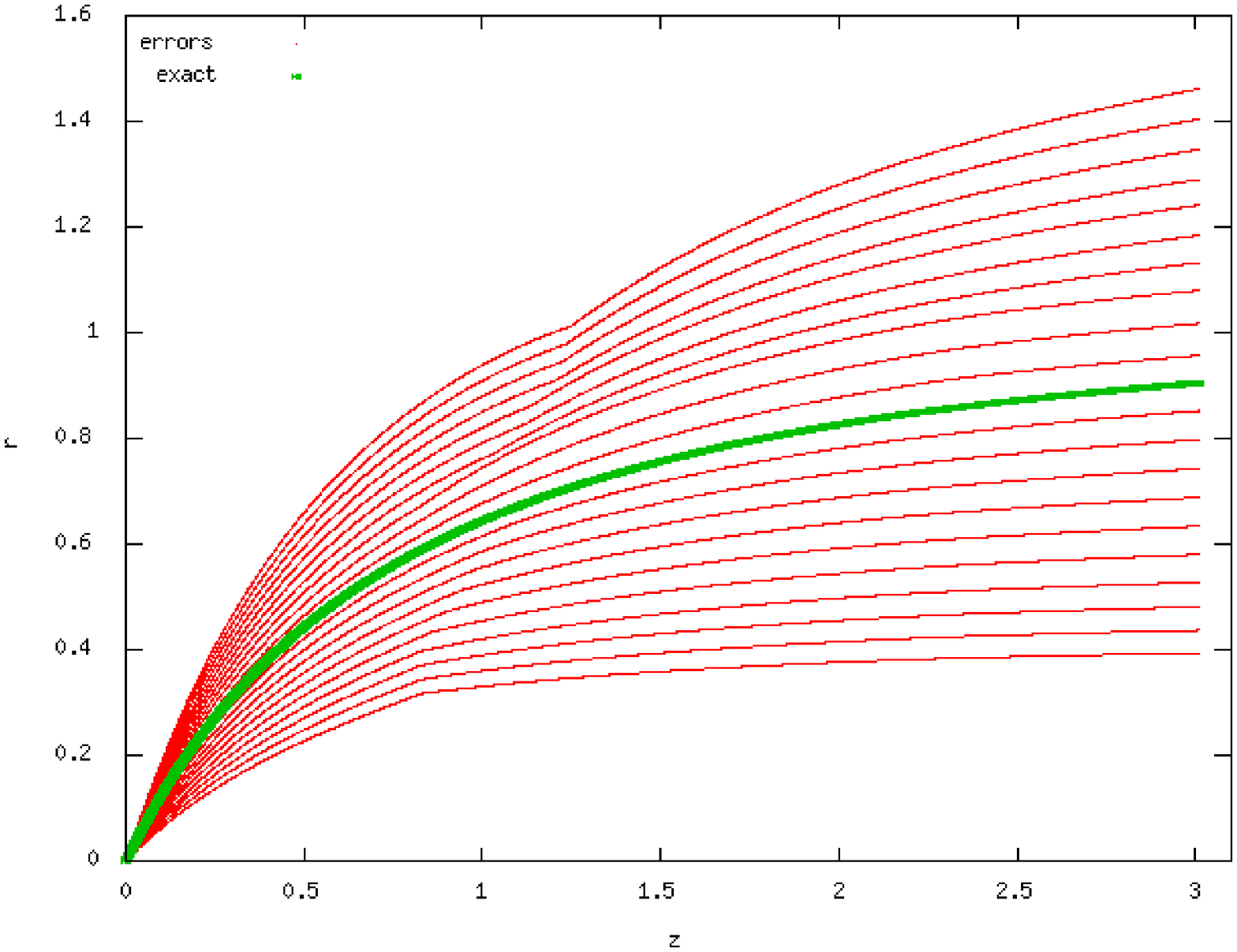}}
\resizebox{88mm}{!}
{\includegraphics{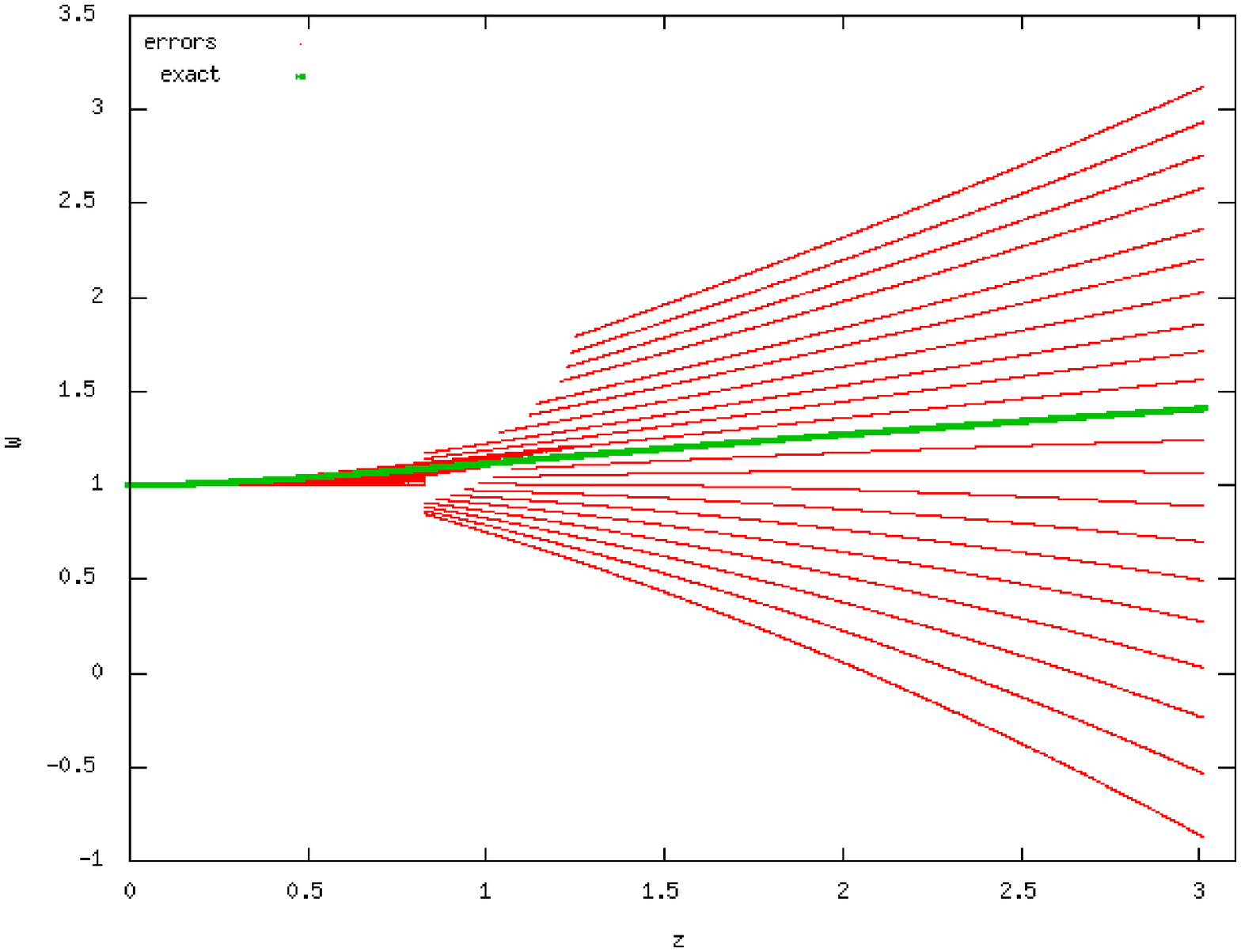}}
\resizebox{88mm}{!}
{\includegraphics{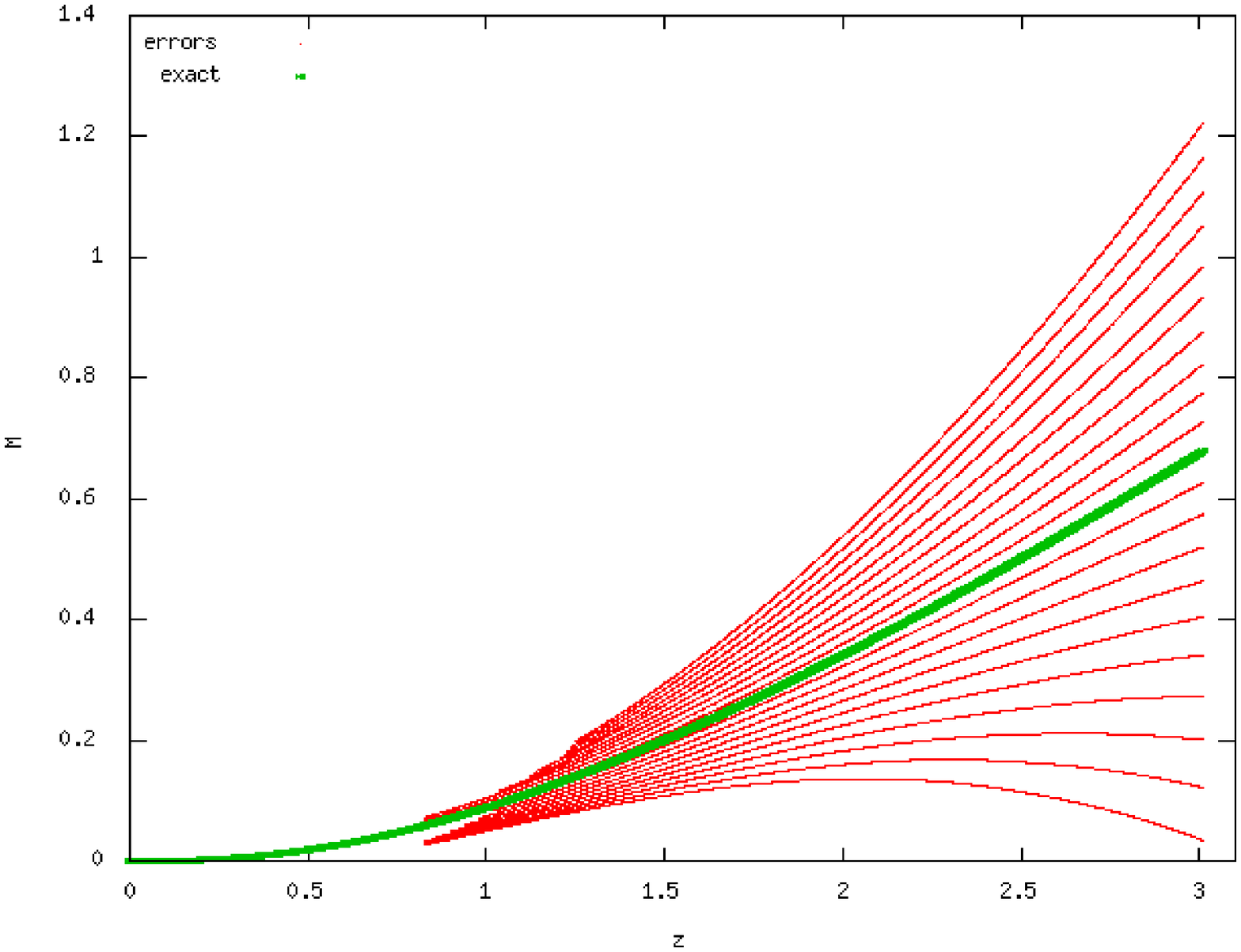}}
\resizebox{88mm}{!}
{\includegraphics{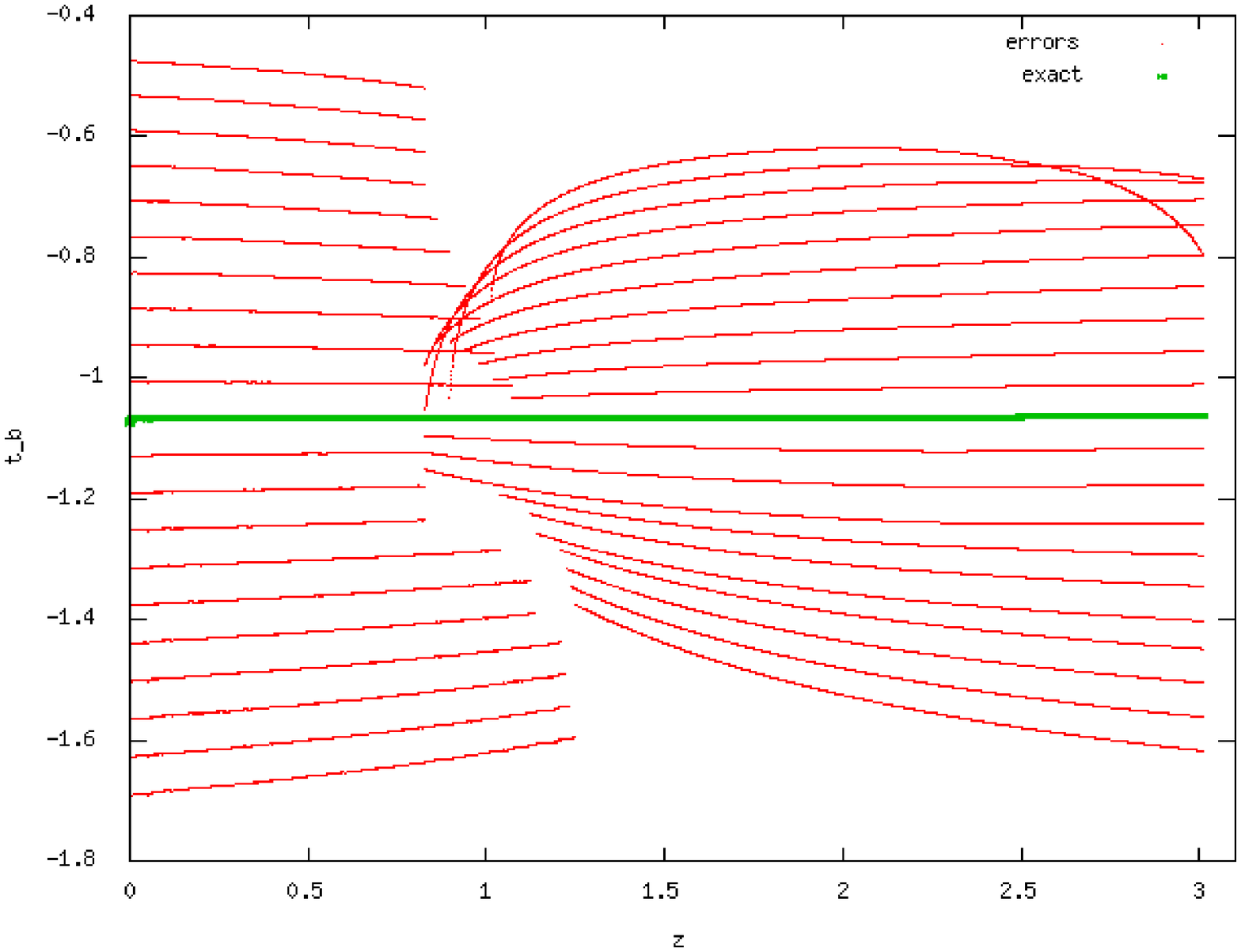}}
\caption{
\figlabel{SystRhErr}
Systematic errors in the data lead to an inconsistency when attempting to
join numerical and near-maximum series regions, leading to jumps in $\phi$,
$W$, $M$ and $t_B$.  This is illustrated for a systematic error in $\hat{R}$
only.  The functions $\phi$ (top left), $r$ (top right), $W$ (middle left),
$M$ (middle right), and $t_B$ (bottom) versus $z$ are plotted with the exact
data (thick line) and with various corrections of the systematic error (thin
lines).  Each subsequent curve represents a 5\% difference in the correction
of the $\Rh$ data.  The plots are for an FLRW universe with $h_0 = 0.72$,
$q_0=0.22$, and $\Omega_{\Lambda}=0$.}
\end{figure*}

It should be noted however that only a constant correction can be made this
way, and a $z$-dependent systematic error cannot be distinguished from a
constant one.  The ability to detect an accumulated systematic error and
correct for it is due to the unique geometric properties of $\Rh_m$ noted by
Hellaby \cite{Hel06}, and this locus is an exception to the theorem of 
Mustapha et al.\ \cite{Mus97}. 

If the systematic percentage error in $\Rh$ grows with $z$, the effect at
low $z$ is qualitatively similar, but less pronounced near the origin. 
However, the maximum in $\Rh$ is changed not only in magnitude $\Rh_m$, but
also in redshift $z_m$.  If this occurs, then the $\phi$ series is calculated
opposingly low (or high) due to the higher (or lower) value of $4 \pi \mu n$
at $z_m$. Then the subsequent numerical integration only asymptotically
corrects the value of $\phi$.  Thus, the overall influence with $\phi$ being
successively too high and too low for parts of the integration is that $r$
may end up approximately correct at the end of the integration.  Both $M$ and
$W$ are not much changed before the series matching, but the shift in $z_m$
opposingly decreases (or increases) the linear slope of the $M$ series curve
more than the high (or low) value in $M_m = \Rh_m/2$ decreases it.  This
makes the values of $M$ and $W$ too low (or high) when the numerical
integration resumes after the near-maximum series expansion.

\subsection{Systematic error in the redshift space density}

With a fixed percentage error in $\mu n$, $\phi$ is initially unaffected, but
eventually becomes correspondingly high (or low) when the $4 \pi \mu n \phi /
(\hat{R} \hat{R}_z)$ term in \er{dphidz2} becomes significant.  The
near-maximum series for $\phi$ is opposingly low (or high) at $\Rh_m$ by
\er{NMSphi0}-\er{NMSphi4}.  Coming out of the series, the subsequent
numerical values of $\phi$ only asymptotically approach the correct value. 
Thus, with $\phi$ being both too high and too low in turns, the integrated
values of $r$ are not as far off by the end of the integration as with a
systematic error in the distance scale.  These effects can be seen in Figure
\ref{Syst4pmnErr}. 

The value of $M$ is initially correspondingly high (or low), since it depends
on the $4 \pi \mu n$ values, but $W$ correspondingly increases (or decreases)
because of the $-M \phi/(\hat{R} \hat{R}_z)$ term. This causes $M$ to be only
marginally high (or low) at the matching, but enough that the linear slope of
the $M$ series (and constant term of the $W$ series) is opposingly low (or
high) so that this then continues throughout the final numerical integration. 
The effects on the $W$ and $M$ values also affects the $t_B$ values.  These
effects can also be seen in Figure \ref{Syst4pmnErr}. 

Again, the presence of a vertical jump in $\phi$ at the matching from
numerical to near-maximum series regions is a distinctive feature of a
systematic error in the data, which may be corrected by adjusting the $\mu n$
data until the jump vanishes. 

The results are largely the same for an error in $\mu n$ that systematically 
grows with redshift, except that obviously the effects are smaller 
if there is no error to begin with and the effects only begin to 
manifest themselves at higher redshift. 

\begin{figure*}
\resizebox{88mm}{!}
{\includegraphics{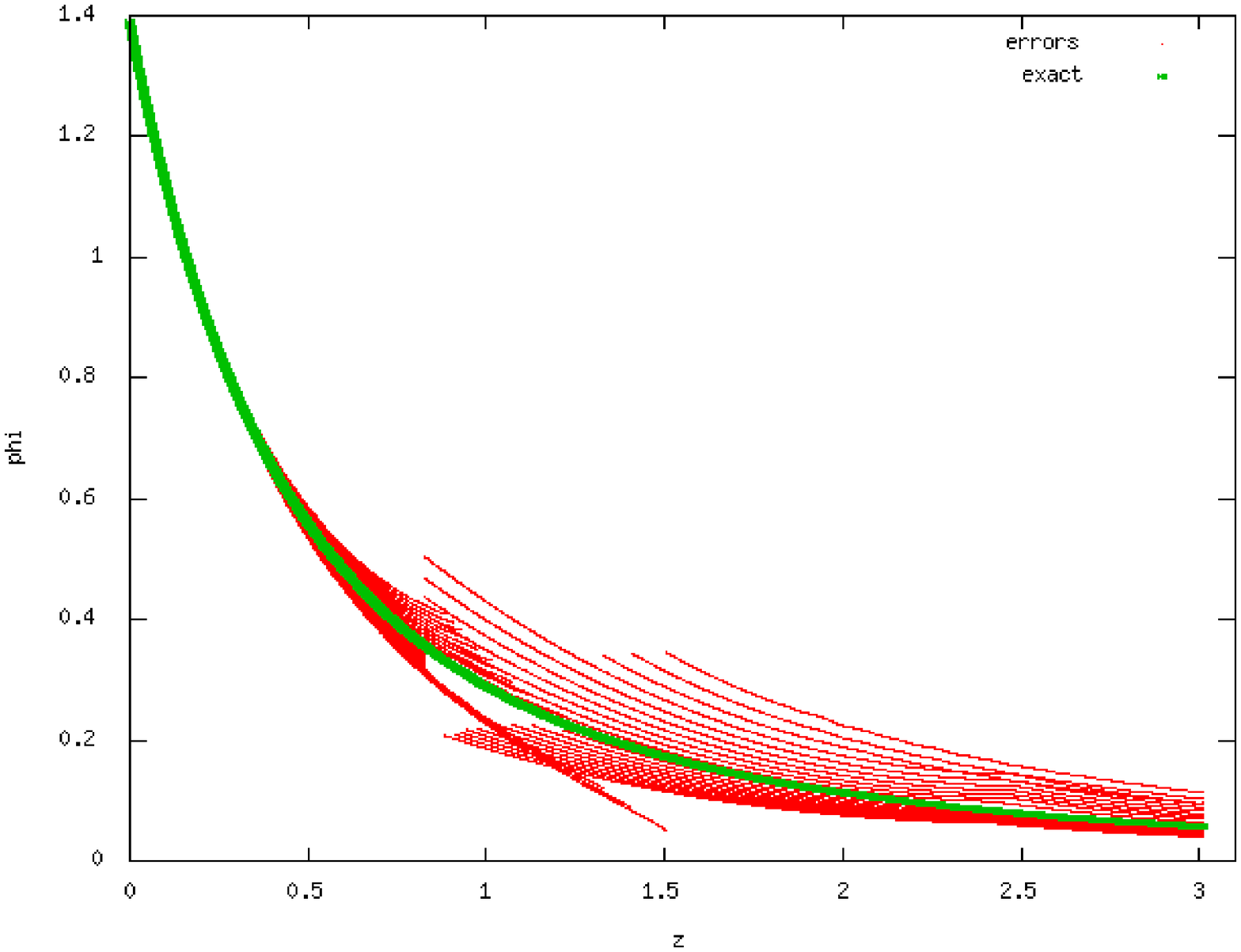}}
\resizebox{88mm}{!}
{\includegraphics{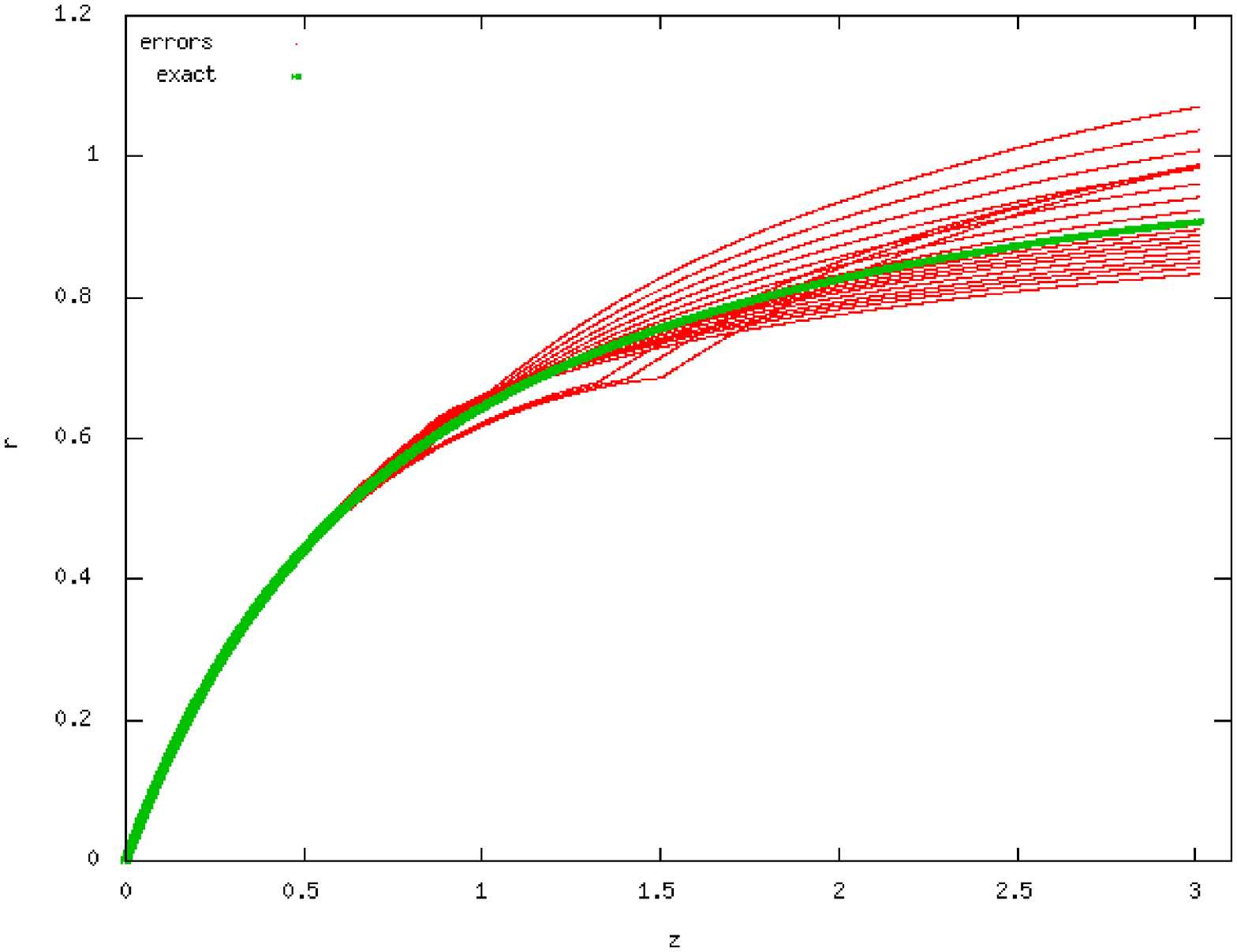}}
\resizebox{88mm}{!}
{\includegraphics{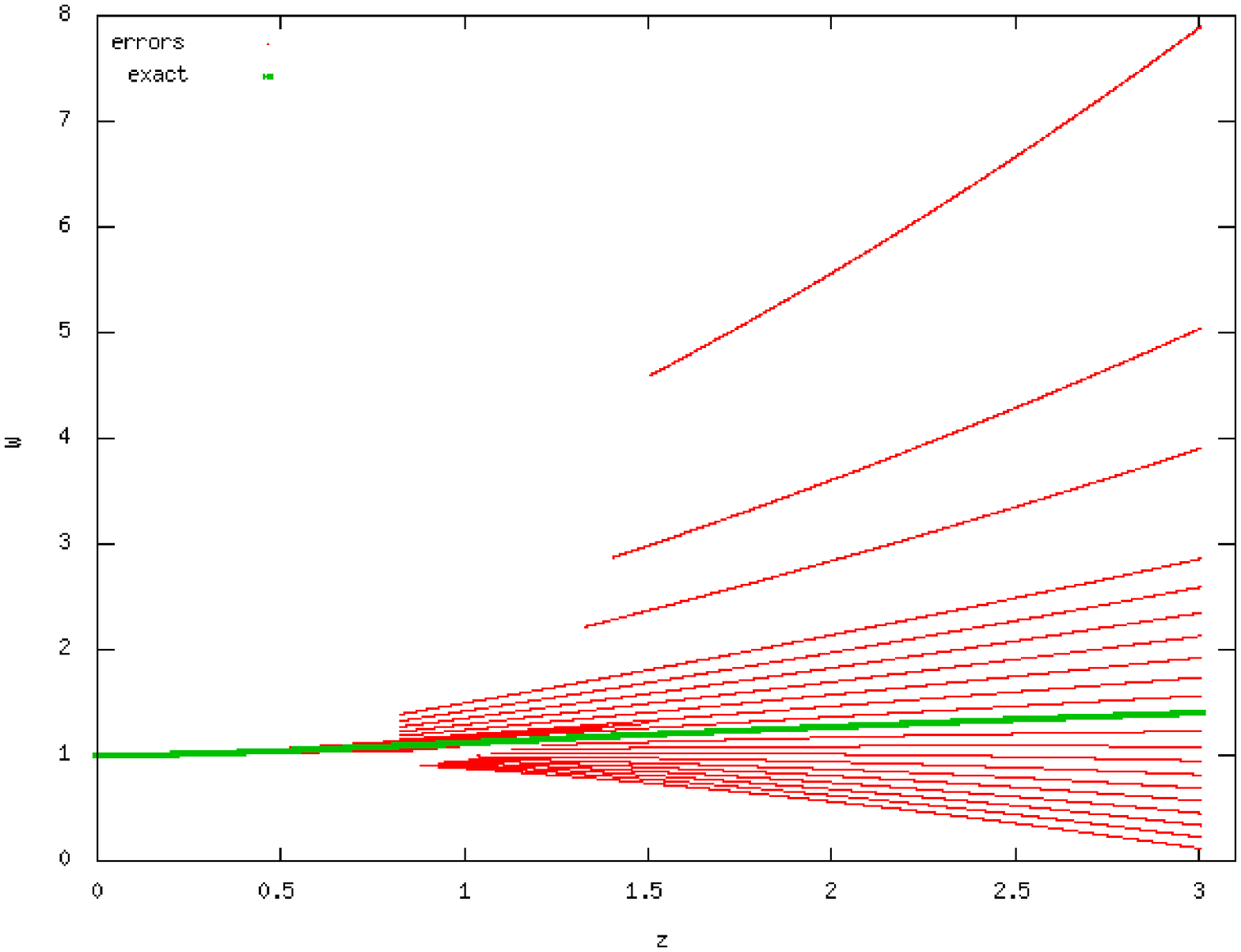}}
\resizebox{88mm}{!}
{\includegraphics{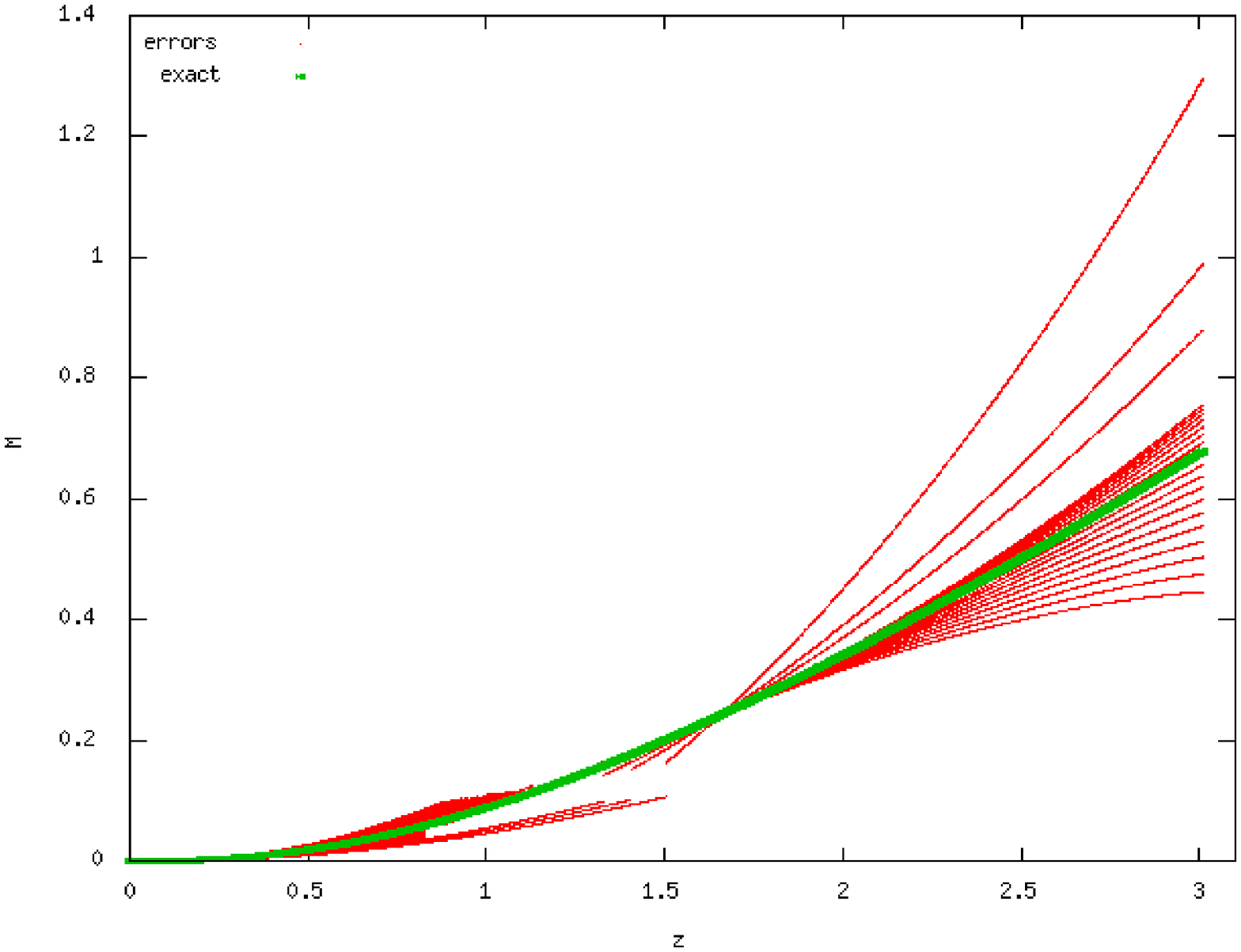}}
\resizebox{88mm}{!}
{\includegraphics{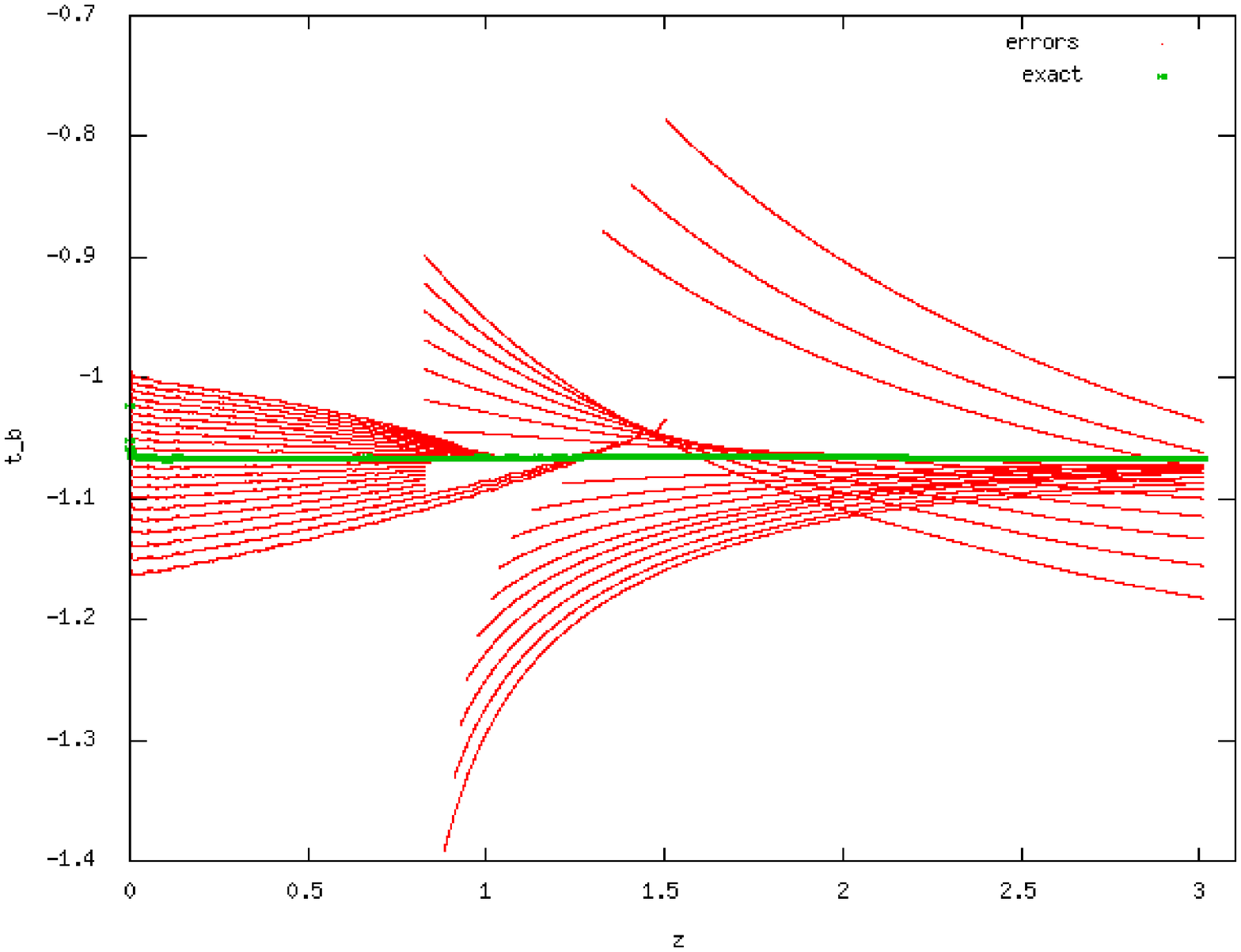}}
\caption{
 \figlabel{Syst4pmnErr}
Similar to Figure \ref{SystRhErr}, but with a systematic error added to the
$4 \pi \mu n$ data instead of the $\hat{R}$ data.  The jumps due to various
corrections of the systematic error in $4 \pi \mu n$ are illustrated for the
functions $\phi$ (top left), $r$ (top right), $W$ (middle left), $M$ (middle
right), and $t_B$ (bottom) versus $z$, with the exact data (thick line) for
comparison.  Each subsequent curve (thin lines) represents a 5\% difference
in the correction of the $4 \pi \mu n$ data.  The plots are for an FLRW
universe with $h_0 = 0.72$, $q_0=0.22$, and $\Omega_{\Lambda}=0$.}
\end{figure*}

\subsection{Systematic errors in both distance and density}

Realistically, we expect both $\Rh$ and $\mu n$ to suffer from systematic
errors.  The plots in Figure \ref{SystBothErrors} show the effects when the
$\hat{R}$ values are all 10\% too large and the $4 \pi \mu n$ values are all
10\% too small.  If a correction is made to the $\hat{R}$ values only, this
leads to an overcorrection such that $r$ and $M$ are 10\% too small and $t_B$
is 10\% too large, compared with the original model the data came from.  If a
correction is made to the $4 \pi \mu n$ values only, this makes $r$ and $M$
10\% too large and $t_B$ 10\% too small.  Once one correction has been made,
there are no jumps left in the data, so it is not apparent any further
correction needs to be made. 

With identical constant percentage errors (say $p$) in both $\hat{R}$ and $4
\pi \mu n$, $\phi$ starts out correspondingly high or low, the $\phi$ series
is only slightly off, and then the $\phi$ integration asymptotically
approaches the correct value at the end.  Thus, $r$ is too low by nearly the
same percentage error, $p$, over the whole run.  The low $4 \pi \mu n$ values
tend to make $M$ low, and the errors in the $-M \phi/(\hat{R} \hat{R}_z)$
term for $W$ essentially cancel out so that $W$ is correct.  Since both the
$M$ matching value and the $M$ series value at the maximum are
correspondingly high or low, then the slope in the $M$ series (and hence the
intercept in the $W$ series) is approximately correct, so $M$ ends up being
off by about $p$ percent over the whole run.  Thus, only in the special case
where both the errors are systematically off by the same percentage in the
same direction is the simple result obtained that $r$ and $M$ are both off by
the same percentage.  In that case, since there is no apparent jump in
$\phi$, there is no way of knowing that an error correction is necessary. 

Therefore, in making corrections for systematic errors, it is only possible
to ensure that the $\hat{R}$ and $\mu n$ data are consistent at the maximum,
i.e.\ that the accumulated mass derived from the integration agrees with that
deduced from $\Rh_m$. 

\begin{figure*}
\resizebox{88mm}{!}
{\includegraphics{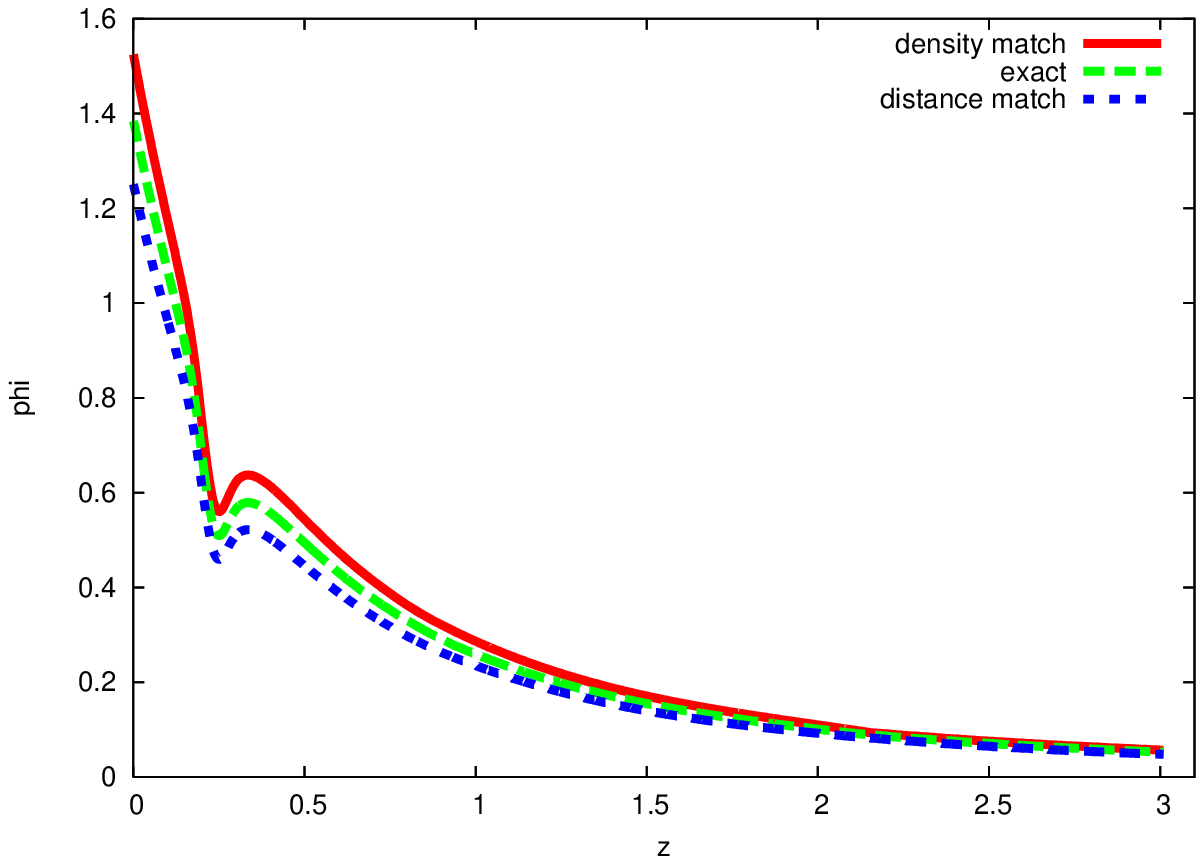}}
\resizebox{88mm}{!}
{\includegraphics{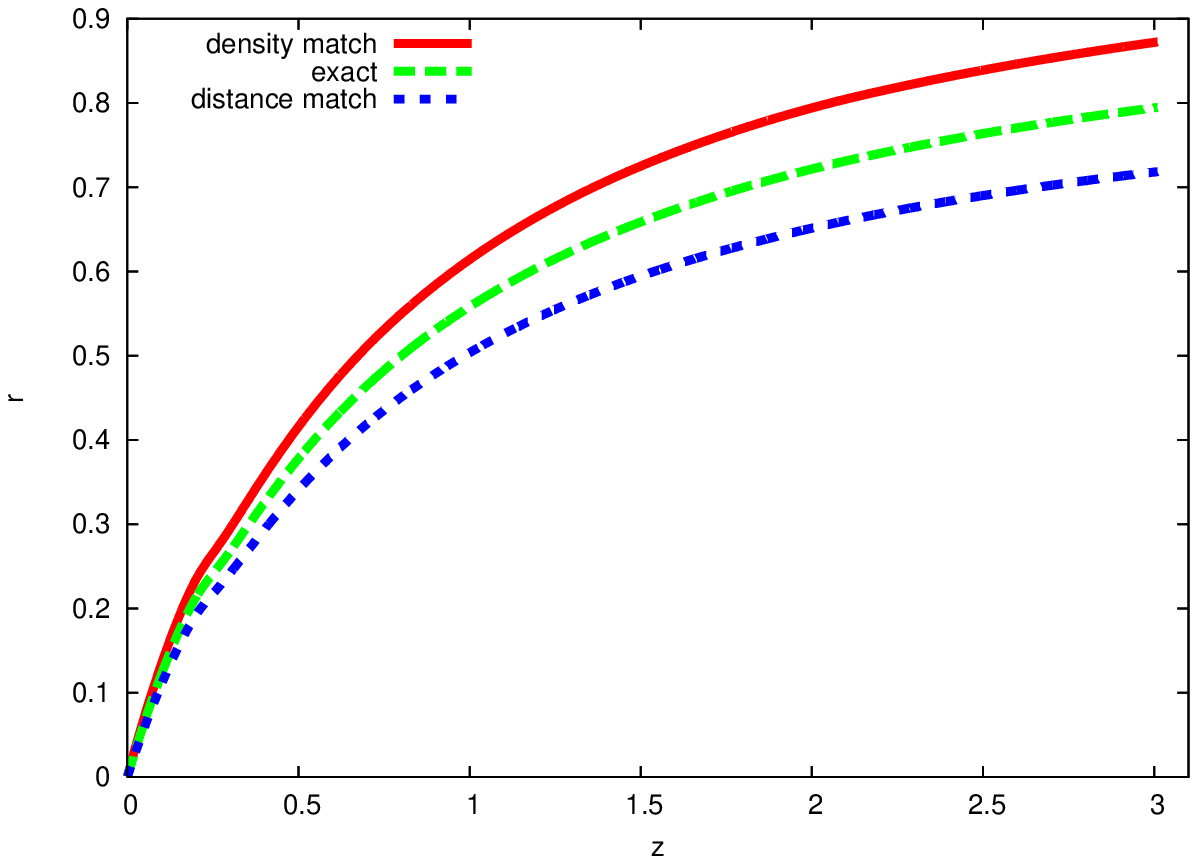}}
\resizebox{88mm}{!}
{\includegraphics{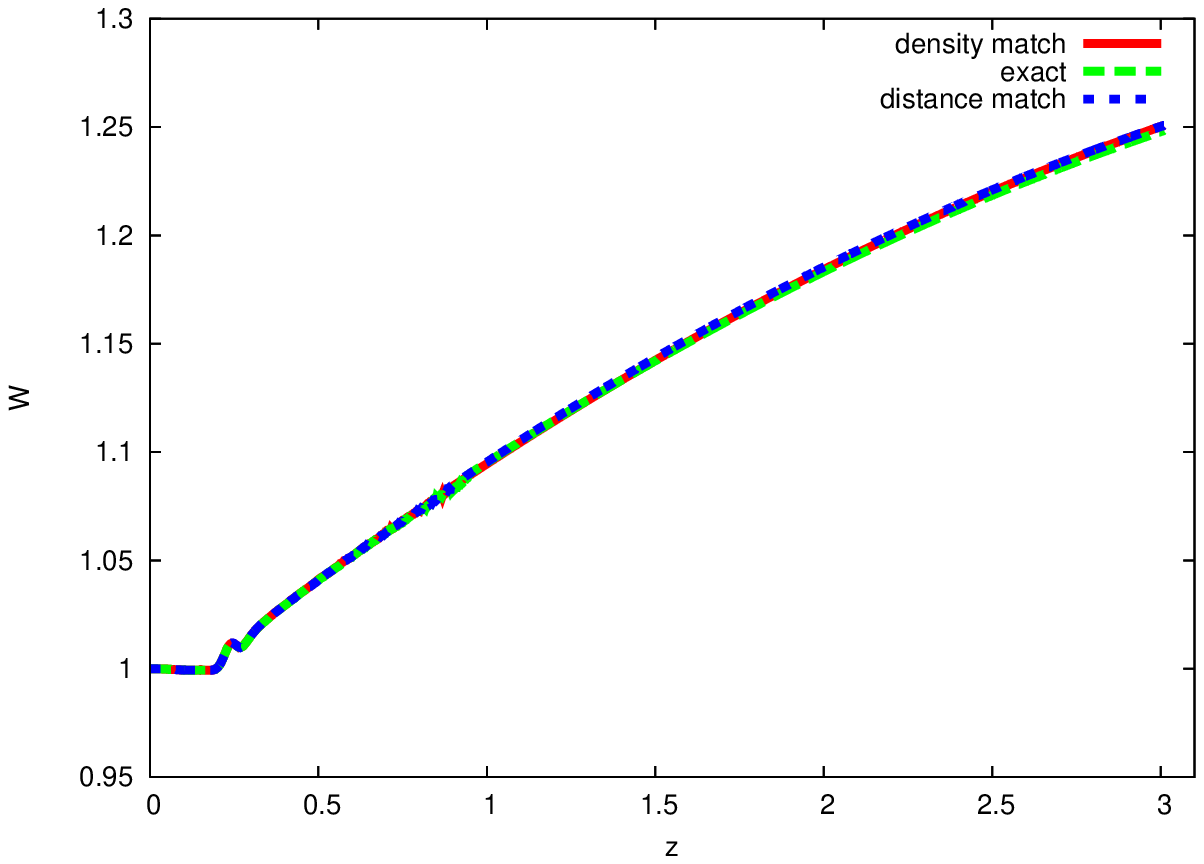}}
\resizebox{88mm}{!}
{\includegraphics{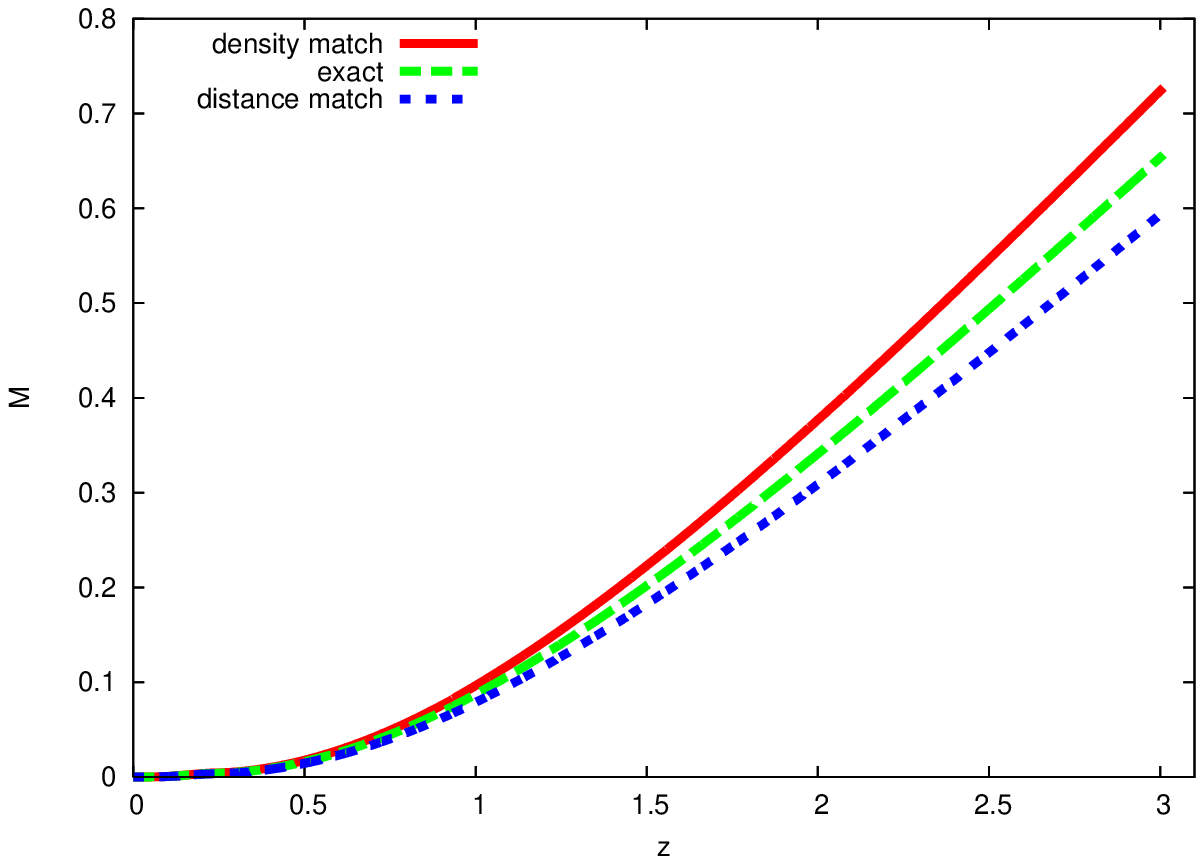}}
\resizebox{88mm}{!}
{\includegraphics{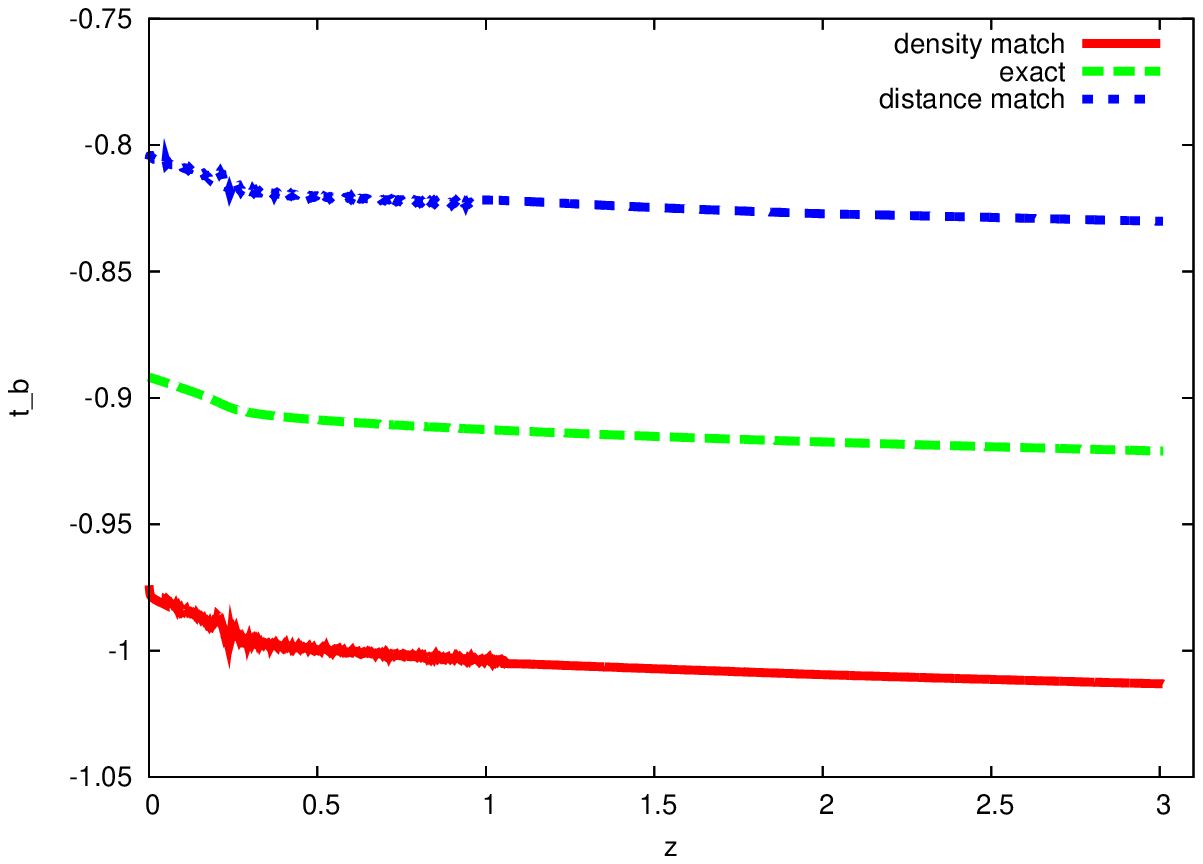}}
\caption{
 \figlabel{SystBothErrors}
Corrected curves with systematic errors added to both the $\Rh$ and $4 \pi
\mu n$ data (+10\% and -10\% respectively).  The plots show the exact data
(dashes) and the corrected curves obtained by matching the curves for the
$\hat{R}$ data (dots) or $4 \pi \mu n$ data (solid curve) to eliminate jumps. 
It is apparent the jumps may be removed by correcting either the $\Rh$ or the
$4 \pi \mu n$ data.  It is not possible to determine which correction should
be made, but either result is much more self-consistent than with no
correction. This is illustrated for the functions $\phi$ (top left), $r$ (top
right), $W$ (middle left), $M$ (middle right), and $t_B$ (bottom) versus $z$. 
The plots of $r$, $M$, and $t_B$ clearly end up overcorrected by 10\% when a
correction is made to one variable.  These plots are for an inhomogeneous
universe with origin parameters $h_0=0.72$, $q_0=0.6$, and
$\Omega_{\Lambda}=0$: a strong inhomogeneity is present around $z=0.2:0.3$.}
\end{figure*}

\section{Conclusion}

The idea of extracting metric information from cosmological observations has
been investigated theoretically in many papers, but until now had never been
turned into a practical procedure.  Such a project began with Lu and 
Hellaby \cite{Lu07} in
which the theoretical algorithm of Mustapha et al.\ \cite{Mus97} was turned 
into a
numerical program that uses galaxy redshifts, luminosity or diameter
distances, and density in redshift space, to determine the geometry of the
Cosmos; that is, it extracts information about the spacetime metric.  The
program was tested with fake data for which the true spacetime metric was
known, and successfully demonstrated the basic viability of the procedure. 
At this early stage of development, spherical symmetry is assumed in order to
focus on issues such as the development of an appropriate numerical method. 
In developing this procedure, it became clear that there are four distinct
integration regions that require different treatment: setting up the initial
values at the origin, a first numerical integration region, the neighbourhood
of the maximum in the diameter distance, and a second numerical region. 

The original code of Lu and Hellaby \cite{Lu07} has been modified to deal 
with statistical and
systematic errors, and we have studied three consequences: how to estimate
uncertainties in the metric functions, how data errors affect the stability
of the integration process, and how a key consistency condition can be used
to identify and partly correct for systematic errors.  Simulated data were
generated by adding Gaussian deviates to ideal data based on various
homogeneous and inhomogeneous model universes, and this was used to test the
new program. Significant improvements were needed in several aspects of the
procedure. Because the DEs to be solved involve $\Rh_{zz}$, the second
derivative of the data function $\Rh$, it is essential to smooth out
observational noise by fitting a polynomial to $\Rh$ over a range of $z$
about each point.  Without this, calculated $R_{zz}$ values would suffer from
gigantic errors.  The estimation of the initial values at the origin was
improved by using a small-$z$ series expansion of the LTB metric, and fitting
to a range of redshift bins near the origin.  Also, the junctions between the
numerical and near-maximum series expansion values for $M$ and $W = \sqrt{1 +
2 E}\;$ were improved so that numerical discrepancies were minimized. 

An important result is the theoretical and numerical investigation of
stability.  The problem of integrating through the maximum in $\Rh$, where
the DEs are singular, was already solved by Lu and Hellaby \cite{Lu07} by 
using a series
expansion about $\Rh_m$.  The numerical integration of $\phi(z)$ was here
shown to be stable away from $\Rh_m$, and the region of serious numerical
instability was avoided by the use of this series.  The numerical integration
of $M$ and $W$, however, is only stable before $\Rh_m$ is reached, and beyond
this point errors in $M$ and $W$ feed back on each other.  Because of this,
it will be difficult to extend this integration far beyond $\Rh_m$ unless
another method can be found.  We applied the quick fix of extending the
near-maximum series for $W$ to larger $z$ values and combining it with the
numerical integration of $M$, which removed the instability and gave
acceptable results for the test models considered up to $z = 3$.  However,
such an approach must necessarily become increasingly inaccurate with $z$. 
We suspect however that this instability is not confined to the LTB model or
spherical symmetry, but is a significant issue for analyzing high-$z$ data,
though the assumption of a homogeneous Robertson-Walker metric may hide the
problem. 

Particularly interesting was the effect of systematic errors.  There is a
relationship \er{RhmEq} (or its generalization \er{LamRhmEq}) between the
mass and the diameter distance, that must hold at the maximum in $\Rh$.  With
data extending out past this maximum, it has been shown that it is possible
to correct for an overall systematic error in either the distance ladder or
the redshift matter density, making them mutually consistent at the maximum. 
Since a correction may be applied to either of the two data functions,
$\Rh(z)$ and $\mu(z) n(z)$, there is a range of possible results.  But the
most likely systematic error will be in the redshift matter density, because
the mass per source will be uncertain due to the fact that this mass has to
include a component for all the material in the Universe that never collapsed
to form objects, and for each source that we do see, there will probably be
many smaller sources that we do not.  Thus, it seems most logical to assume
the distance ladder is correct and make the correction in the redshift matter
density. 

Our broad conclusion is that when sufficient cosmological data become
available out towards a redshift $z \sim 2$, it should be possible to
calculate the metric for the Cosmos, and thereby check homogeneity, at least
within this range.  With upcoming surveys, such data should be available in
the not-too-distant future. At larger redshift values, well beyond $\Rh_m$ at
roughly $z \sim 1.5$, there is a difficulty with stability, potentially
resulting in large errors, which we believe may be a general problem for
analysis of high-$z$ data, making it difficult to properly verify homogeneity
on the largest scales.  It is possible that another method might be devised
to avoid this instability, but we have not found one. 

\begin{acknowledgments}

We thank Byron Desnoyers Winmill for programming advice. MLM thanks South
Africa's National Research Foundation for a postdoctoral fellowship.  CH
thanks the National Research Foundation for a research grant. 

\end{acknowledgments}

 \appendix

 \section{The Near-Maximum Series for the PNC in LTB Models}
 \seclabel{NearMaxSeries}

 This appendix provides the solution for the LTB arbitrary functions as series in
$\Delta z = z - z_m$ near $z = z_m$ where the maximum $R_m = \Rh(z_m)$ occurs.  
We write out all functions as power series in $\Delta z$:
 \begin{align}
   \Rh & = R_m + \sum_{i=2}^\infty R_i \Delta z^i ~,~~~~~~~~
      4 \pi \mu n = K_m + \sum_{i=1}^\infty K_i \Delta z^i ~,~~~~~~~~
      \rh = r_m + \sum_{i=1}^\infty r_i \Delta z^i ~, \\
   M & = M_m + \sum_{i=1}^\infty M_i \Delta z^i ~,~~~~~~~~
      \sqrt{1 + 2 E}\; = W = W_m + \sum_{i=1}^\infty W_i \Delta z^i ~.
 \end{align}
 The coefficients in the $\Rh$ and $4 \pi \mu n$ series are found from polynomial 
fits to the data near the maximum in $\Rh$, and the coefficients of the series for 
$\rh$, $M$, and $W$ are obtained by substituting these series into the DEs \er{phiDef}, 
\er{dphidz}, \er{dMdz}, and \er{WPNCz}.  The following extends the results of 
Lu and Hellaby \cite{Lu07}.
We find
 \begin{align}
   \phi_0 = r_1 & = \frac{- 2 R_m R_2}{K_m} ~,   \label{NMSphi0} \\
   \phi_1 = r_2 & = \Bigg( \Bigg\{ \frac{K_1}{K_m} - \frac{1}{1 + z_m} \Bigg\} R_2 - 3 R_3 \Bigg)
      \frac{R_m}{K_m} ~, \\
   \phi_2 = r_3 & = \Bigg( \Bigg\{ \frac{2 K_2}{3 K_m} - \frac{K_1^2}{2 K_m^2}
         + \frac{2 K_1}{3 K_m (1 + z_m)} + \frac{1}{2 (1 + z_m)^2} \Bigg\} R_2 \nn \\
      &~~~~ + \Bigg\{ \frac{K_1}{K_m} - \frac{1}{(1 + z_m)} \Bigg\} \frac{3 R_3}{2} - 4R_{4}
         - \frac{2 R_2^2}{3 R_m} \Bigg) \frac{R_m}{K_m} ~, \\
   \phi_3 = r_4 & = \Bigg( \Bigg\{ \frac{K_3}{2 K_m} - \frac{2 K_1 K_2}{3 K_m^2}
         + \frac{K_1^3}{4 K_m^3} + \frac{K_2}{2 K_m (1 + z_m)} \nn \\
      &~~~~ - \frac{5 K_1^2}{12 K_m^2 (1 + z_m)} - \frac{K_1}{4 K_m (1 + z_m)^2}
         - \frac{1}{4 (1 + z_m)^3} \Bigg\} R_2 \nn \\
      &~~~~ + \Bigg\{ \frac{K_2}{K_m} - \frac{3 K_1^2}{4 K_m^2}
         + \frac{K_1}{K_m (1 + z_m)} + \frac{3}{4 (1 + z_m)^2} \Bigg\} R_3 \nn \\
      &~~~~ + \Bigg\{ \frac{K_1}{K_m} - \frac{1}{1 + z_m} \Bigg\} (2 R_4) - 5 R_5 \nn \\
      &~~~~ + \Bigg\{ \frac{K_1}{6 K_m} - \frac{1}{2 (1 + z_m)} \Bigg\} \frac{R_2^2}{R_m}
         - \frac{3 R_2 R_3}{2 R_m}
         \Bigg) \frac{R_m}{K_m} ~, \\
   \phi_4 = r_5 & = \Bigg( \Bigg\{ \frac{2 K_4}{5 K_m} - \frac{K_1 K_3}{2 K_m^2} - \frac{2 K_2^2}{9 K_m^2}
         + \frac{K_2 K_1^2}{2 K_m^3} - \frac{K_1^4}{8 K_m^4} + \frac{2 K_3}{5 K_m (1 + z_m)} \nn \\
      &~~~~ - \frac{11 K_1 K_2}{18 K_m^2 (1 + z_m)}
         + \frac{K_1^3}{4 K_m^3 (1 + z_m)} + \frac{K_1^2}{9 K_m^2 (1 + z_m)^2} \nn \\
      &~~~~ - \frac{K_2}{6 K_m (1 + z_m)^2} + \frac{K_1}{12 K_m (1 + z_m)^3}
         + \frac{1}{8 (1 + z_m)^4} \Bigg\} R_2 \nn \\
      &~~~~ + \Bigg\{ \frac{3 K_3}{4 K_m} - \frac{K_1 K_2}{K_m^2} + \frac{3 K_1^3}{8 K_m^3}
         + \frac{3 K_2}{4 K_m (1 + z_m)} \nn \\
      &~~~~ - \frac{5 K_1^2}{8 K_m^2 (1 + z_m)} - \frac{3 K_1}{8 K_m (1 + z_m)^2}
         - \frac{3}{8 (1 + z_m)^3} \Bigg\} R_3 \nn \\
      &~~~~ + \Bigg\{ \frac{4 K_2}{3 K_m} - \frac{K_1^2}{K_m^2} + \frac{4 K_1}{3 K_m (1 + z_m)}
         + \frac{1}{(1 + z_m)^2} \Bigg\} R_4 \nn \\
      &~~~~ + \Bigg\{ \frac{K_1}{K_m} - \frac{1}{(1 + z_m)} \Bigg\} \frac{5 R_5}{2}
         - 6 R_6 \nn \\
      &~~~~ + \Bigg\{ \frac{2 K_2}{45 K_m} + \frac{19 K_1}{90 K_m (1 + z_m)}
         + \frac{1}{6 (1 + z_m)^2} \Bigg\} \frac{R_2^2}{R_m} \nn \\
      &~~~~ + \Bigg\{ \frac{7 K_1}{20 K_m} - \frac{23 }{20 (1 + z_m)} \Bigg\} \frac{R_2 R_3}{R_m} \nn \\
      &~~~~ - \frac{3 R_3^2}{4 R_m} - \frac{26 R_2 R_4}{15 R_m}  + \frac{8 R_2^3}{45 R_m^2}   \label{NMSphi4}
 \Bigg) \frac{R_m}{K_m}
 \end{align}
 \begin{align}
   M_m & = \frac{R_m}{2} ~, \\
   M_1 & = M_1 ~, \\
   M_2 & = \Bigg\{ \frac{K_1}{K_m} + \frac{1}{1 + z_m} \Bigg\} \frac{M_1}{2} 
      - \frac{R_2}{2} - \frac{K_m^2}{2 R_m} ~, \\
   M_3 & = \Bigg\{ \frac{K_2}{K_m} + \frac{K_1}{K_m (1 + z_m)}
         - \frac{R_2}{R_m} \Bigg\} \frac{M_1}{3} \nn \\
      &~~~~ - \Bigg\{ \frac{K_1}{K_m} + \frac{1}{1 + z_m} \Bigg\} \frac{R_2}{4}
         - \frac{R_3}{4} - \frac{K_m K_1}{2 R_m} ~, \\
   M_4 & = \Bigg\{ \frac{K_3}{K_m} + \frac{K_2}{K_m (1 + z_m)} - \frac{K_1 R_2}{K_m R_m}
         - \frac{R_3}{R_m} - \frac{1}{R_m (1 + z_m)} \Bigg\} \frac{M_1}{4} \nn \\
      &~~~~ - \Bigg\{ \frac{5 K_1}{36 K_m (1 + z_m)} + \frac{2 K_2}{9 K_m} - \frac{K_1^2}{24 K_m^2}
         - \frac{K_m^2}{6 R_m^2} - \frac{1}{24 (1 + z_m)^2} \Bigg\} R_2
         + \frac{2 R_2^2}{9 R_m} \nn \\
      &~~~~ - \Bigg\{ \frac{K_1}{8 K_m} + \frac{1}{8 (1 + z_m)} \Bigg\} R_3
 - \frac{R_4}{6} - \frac{K_1^2}{8 R_m} - \frac{K_2 K_m}{3 R_m} - \frac{K_m^2}{24 R_m (1 + z_m)^2} ~;
 \end{align}
 and 
 \begin{align}
   W_m & = \frac{M_1}{K_m} ~, \\
   W_1 & = \frac{M_1}{K_m (1 + z_m)} - \frac{R_2}{K_m} - \frac{K_m}{R_m} ~, \\
   W_2 & = - \frac{R_2 M_1}{R_m K_m} + \Bigg\{ \frac{K_1}{4 K_m} - \frac{3}{4 (1 + z_m)}
         \Bigg\} \frac{R_2}{K_m} \nn \\
      &~~~~ - \frac{3 R_3}{4 K_m} - \frac{K_1}{2 R_m} ~, \\
   W_3 & = - \Bigg\{ R_3 + \frac{R_2}{(1 + z_m)} \Bigg\} \frac{M_1}{R_m K_m} \nn \\
      &~~~~ + \Bigg\{ \frac{K_2}{9 K_m} - \frac{K_1^2}{12 K_m^2} + \frac{2 K_m^2}{3 R_m^2}
         + \frac{7 K_1}{36 K_m (1 + z_m)} \nn \\
      &~~~~ + \frac{1}{6 (1 + z_m)^2} \Bigg\} \frac{R_2}{K_m} + \frac{8 R_2^2}{9 R_m K_m} \nn \\
      &~~~~ + \Bigg\{ \frac{K_1}{4 K_m} - \frac{1}{2 (1 + z_m)} \Bigg\} \frac{R_3}{K_m}
            - \frac{2 R_4}{3 K_m} \nn \\
      &~~~~ - \frac{K_2}{3 R_m} - \frac{K_m}{6 R_m (1 + z_m)^2} ~.
 \end{align}


\begin{thebibliography}{99}

\def\cqg{{Class.\ Quantum Grav.\/} }
\def\grg{{Gen.\ Relativ.\ Gravit.\/} }
\def\pm7{{Philos.\ Mag.\ VII} }
\def\prs{{Proc.\ R. Soc.\ London} }
\def\zfa{{Z. Astrophys.} }

\bibitem{Tem38} G. Temple, \prs {\bf A168}, 122 (1938).
\bibitem{McC34} W. H. McCrea, \zfa {\bf 9}, 290 (1934).
\bibitem{McC38} W. H. McCrea, \zfa {\bf 18}, 98 (1939) [\grg {\bf 30}, 
315 (1998)].
\bibitem{Kri66} J. Kristian and R. K. Sachs, Astrophys.\ J. \textbf{143}, 
379 (1966).
\bibitem{Ell85} G. F. R. Ellis, S. D. Nel, R. Maartens, W. R. Stoeger, 
and A. P.
Whitman, Phys.\ Reports. \textbf{124}, 315 (1985).
\bibitem{AS99} M. E. Ara\'{u}jo and W. R. Stoeger, \prd {\bf 60}, 104020 
(1999)
[erratum: \prd {\bf 64}, 049902 (2001)].
\bibitem{AABFS01} M. E. Ara\'{u}jo, R. C. Arcuri, M. L. Bedran, L. R. de 
Freitas,
and W. R. Stoeger, \apj {\bf 549}, 716 (2001).
\bibitem{ARS01} M. E. Ara\'{u}jo, S. R. M. M. Roveda, and W. R. 
Stoeger, \apj
{\bf 560}, 7 (2001).
\bibitem{MaartensMatravers} R. Maartens and D. R. Matravers, \cqg {\bf 
11}, 2693
(1994).
\bibitem{MHMS96} R. Maartens, N. P. Humphreys, D. R. Matravers, and W. R.
Stoeger, \cqg {\bf 13}, 253 (1996) [erratum: \cqg {\bf 13}, 1689 (1996)].
\bibitem{SNME92} W. R. Stoeger, S. D. Nel, R. Maarteens, and G. F. R. Ellis,
\cqg {\bf 9}, 493 (1992).
\bibitem{SEN92} W. R. Stoeger, G. F. R. Ellis, and S. D. Nel, \cqg {\bf 
9}, 509
(1992).
\bibitem{SNE92b} W. R. Stoeger, S. D. Nel, and G. F. R. Ellis,
\cqg {\bf 9}, 1711 (1992).
\bibitem{SNE92c} W. R. Stoeger, S. D. Nel, and G. F. R. Ellis, \cqg {\bf 9},
1725 (1992).
\bibitem{MBHE} N. Mustapha, B. A. C. C. Bassett, C. Hellaby, and G. F. R. 
Ellis, \cqg {\bf 15}, 2363 (1998).
\bibitem{RS03} M. B. Ribeiro and W. R. Stoeger, \apj {\bf 592}, 1 (2003).
\bibitem{AlIrRiSt07} V. V. Albani, A. S. Iribarrem, M. B. Ribeiro, and W. R. 
Stoeger, \apj {\bf 657}, 760 (2007).
\bibitem{Mus97} N. Mustapha, C. W. Hellaby, and G. F. R. Ellis, Mon.\
Not.\ R.\ Astron.\ Soc. \textbf{292}, 817 (1997).
\bibitem{Lem33} G. Lema\^{i}tre, Ann.\ Soc.\ Sci. Bruxelles.
\textbf{A53}, 51 (1933) [Gen.\ Relativ.\ Gravit. \textbf{29}, 641 (1997)].
\bibitem{Tol34} R. C. Tolman, Proc.\ Nat.\ Acad.\ Sci.\ U.S.A. \textbf{20},
169 (1934) [Gen.\ Relativ.\ Gravit. \textbf{29}, 935 (1997)].
\bibitem{Bon47} H. Bondi, Mon.\ Not.\ R.\ Astron.\ Soc. \textbf{107}, 410
(1947) [Gen.\ Relativ.\ Gravit. \textbf{11}, 1783 (1999)].
\bibitem{Hel01} C. Hellaby, Astron.\ Astrophys.\ \textbf{372}, 357 (2001).
\bibitem{Cel00} M.-N. C\'el\'erier, Astron.\ Astrophys.\ {\bf 353}, 63 
(2000).
\bibitem{Cel07a} M.-N. C\'el\'erier, New Adv.\ Phys.\ {\bf 1}, 29 (2007).
\bibitem{Cel07b} M.-N. C\'el\'erier, \verb"arXiv:0706.1029", to appear in
Proceedings of the XIXth Rencontres de Blois, Matter and energy in the 
Universe, Blois, France, May 2007.
\bibitem{Bish96} N. Bishop and P. Haines, Quaestiones Mathematicae. 
\textbf{19}, 259 (1996).
\bibitem{Lu06} H.-C. Lu, M.Sc.\ Thesis, University of Cape Town, 2006.
\bibitem{Lu07} T. H.-C. Lu and C. Hellaby, \cqg {\bf 24}, 4107 (2007).
\bibitem{Hel06} C. Hellaby, Mon.\ Not.\ R.\ Astron.\ Soc. \textbf{370}, 239
(2006).
\bibitem{Ellis71} G. F. R. Ellis, in {\it General Relativity and
Cosmology: Proc. Int. School Phys. ``Enrico Fermi'' (Varenna) Course 
XLVII}, edited by R. K. Sachs (Academic Press, New York, 1971), p.\ 104.
\bibitem{Eth33} I. M. H. Etherington, \pm7 {\bf 15}, 761 (1933)
 [\grg {\bf 39}, 1055 (2007)].
\bibitem{MH01} N. Mustapha and C. Hellaby, \grg {\bf 33}, 455 (2001).
\bibitem{KraHe04} A. Krasi\'{n}ski and C. Hellaby, \prd {\bf 69}, 043502 
(2004).
\bibitem{H87} C. Hellaby, \cqg {\bf 4}, 635 (1987).
\bibitem{Pre02} W. H. Press, S. A. Teukolsky, W. T. Vetterling, and 
B. P. Flannery, \textit{Numerical recipes in C++: the art of 
scientific computing}, 2nd ed.\ (Cambridge University Press, Cambridge, 
2002), p.\ 294.  

\end{thebibliography}
\end{document}